# *Zero-Field J-spectroscopy of Quadrupolar Nuclei*

Román Picazo-Frutos,[a] Kirill F. Sheberstov,[a,b] John W. Blanchard,[a,c] Erik Van Dyke,[a]

Moritz Reh,[d,e] Tobias Sjoelander,[f,g] Alexander Pines,[g]

Dmitry Budker,[a,d] and Danila A. Barskiy*,[a,g]

[a]Johannes Gutenberg-Universitat Mainz, 55128 Mainz, Germany

Helmholtz-Institut Mainz, GSI Helmholtzzentrum für Schwerionenforschung GmbH, 55128 Mainz, Germany

[b]Department of Chemistry, École Normale Supérieure, PSL University, Paris, France

[c]Quantum Technology Center, University of Maryland, College Park, MD, USA

[d]Department of Physics, University of California – Berkeley, Berkeley, California 94720, USA

[e]Kirchhoff-Institut für Physik, Universität Heidelberg, Im Neuenheimer Feld 227, 69120 Heidelberg, Germany

[f]Department of Physics, University of Basel, Klingelbergstrasse 82, Basel, CH-4056, Switzerland

[g]Department of Chemistry, University of California - Berkeley, California 94720-3220, USA

Materials Science Division, Lawrence Berkeley National Laboratory, Berkeley, California 94720-3220, USA

*Corresponding author: dbarskiy@uni-mainz.de (Danila A. Barskiy)

**ABSTRACT**

Zero- to ultralow-field nuclear magnetic resonance (ZULF NMR) is a powerful version of NMR that allows studying molecules and their transformations in the regime dominated by intrinsic spin-spin interactions. While spin dynamics at zero magnetic field can be probed indirectly—via shuttling a sample that underwent evolution at zero field to a high-field NMR spectrometer for detection—*J*-spectra can also be measured directly at zero field by using non-inductive sensors, for example, atomic magnetometers. To date, no zero-field *J*-spectra of molecules featuring the coupling to quadrupolar nuclei were reported. Here we show that zero-field *J*-spectra can be collected from molecules containing quadrupolar nuclei with $I$ = 1 and demonstrate this for solutions containing various isotopologues of ammonium cations, namely, $^{14}\mathrm{NH}_4^+$ and $^{15}\mathrm{ND}_x\mathrm{H}_{4-x}^+$ (where $x$ = 0, 1, 2, or 3). Lower ZULF NMR signals are observed for molecules containing larger numbers of deuterons compared to protons; this is attributed to less overall magnetization and not to the scalar relaxation of the second kind. Values for the $^{15}$N-$^{1}$H and $^{14}$N-$^{1}$H *J*-couplings of -73.416(3) Hz, 52.395(2) Hz, respectively, are extracted from the ZULF NMR spectra. Precision measurement of the $\left|J_{^{15}\mathrm{NH}}/J_{^{14}\mathrm{NH}}\right|$ ratio resulted in the value within a range of $1.4009 - 1.4013$ depending on specific *J*-peaks used for analysis; this is statistically different from $\left|\gamma_{^{15}\mathrm{N}}/\gamma_{^{14}\mathrm{N}}\right| = 1.4027$ reported in the literature indicating the presence of spin interactions unaccounted for by a simple model. We analyze the energy structure for the studied molecular cations in detail and demonstrate that spectral line positions depend dramatically both on the sign of each *J*-coupling and on the magnetic pulse length. Simple symmetric cations containing quadrupolar nuclei do not require extensive isotopic labeling for the observation of *J*-spectra and, thus, in combination with ZULF NMR may find applications in biomedicine and energy storage.

# 1. Introduction

Zero- to ultralow-field nuclear magnetic resonance (ZULF NMR) is a variant of NMR in which measurements are performed in the absence of a large external magnetic field.[1-2] In such a regime (as opposed to conventional high-field NMR), intrinsic spin-spin interactions—*J*-couplings and dipole-dipole couplings—are not truncated by the coupling to the external magnetic field.[3-5] This condition opens a way for obtaining unique chemical information with modest instrumentation costs.[6-7] Because of its ability to detect subtle intrinsic spin-spin interactions that provide valuable information about chemical composition of the sample under study, ZULF NMR can serve as an advantageous detection modality in situations where the use of expensive superconducting magnets is undesirable or impossible.[8] In particular, the analysis of biologically-relevant samples (e.g., natural extracts or metabolites in bioreactors) using portable ZULF NMR sensors is a desirable goal.[9-10] In addition, applications of ZULF NMR include the search for axion-like dark matter and tabletop studies of physics beyond the standard model.[11]

A typical ZULF NMR experiment consists of the following steps: (i) pre-polarizing a sample in an external magnetic field, (ii) shuttling of the sample to the zero-magnetic-field region, (iii) applying magnetic field pulse(s) to the sample (or fast, non-adiabatic switching off of the guiding magnetic field) to generate a coherent nuclear spin evolution at zero field, and (iv) acquiring the NMR signal using a sensitive (5-10 $fT/Hz^{1/2}$) atomic magnetometer.[5-6] Processing of the signal is performed in a manner similar to high-field NMR experiments (i.e., via Fourier transformation).[12]

The first ZULF NMR measurements were performed on a handful of simple molecules containing a small number (2-5) of *J*-coupled nuclear spins. It was noted that in some cases ZULF NMR measurements did not result in observable *J*-spectra for substances under study.[13] Systems that do not produce observable ZULF NMR spectra typically contain spin-1 nuclei, such as deuterium (D), $^{14}$N, or $^{35}$Cl.[14-16] These nuclei are quadrupolar since, besides having a magnetic dipole moment, they possess electric quadrupole moment and, therefore, interact with the electric field gradients. For this reason, quadrupolar coupling typically dominates other nuclear spin interactions and causes, in the presence of molecular tumbling, fast relaxation on the order of few milliseconds (compared to seconds for spin-1/2 nuclei) leading to broadening of the ZULF NMR lines.[12, 17-18]

In this work, we systematically study the effect of quadrupolar nuclei on the observable ZULF NMR *J*-spectra. As a system for such study, we have chosen aqueous solutions of ammonium cations prepared in mixtures of $H_2O$ and $D_2O$ at high acidity (concentration of $H_2SO_4$ is ~2 M). We first demonstrate a *J*-spectra of [$^{14}$N]-ammonium ($^{14}\mathrm{NH}_4^+$) and [$^{15}$N]-ammonium ($^{15}\mathrm{NH}_4^+$) by taking advantage of the unique tetrahedral environment of the hydrogen atoms which effectively switches off quadrupolar interactions of the $^{14}$N nucleus (**Figure 1**). Expensive isotopic labeling is not required for the preparation of these spin systems; this extends application opportunities for symmetric molecular ions in zero-field *J*-spectroscopy. We then report the most precise to date measurement of the $\left|J_{^{15}\mathrm{NH}}/J_{^{14}\mathrm{NH}}\right|$ ratio which we found to be in the range of $1.4009-1.4013$ (depending on specific peaks used for analysis) which is statistically different from $\left|\gamma_{^{15}\mathrm{N}}/\gamma_{^{14}\mathrm{N}}\right| = 1.4027$ listed in the table data for gyromagnetic ratios. Precision measurements of *J*-couplings performed on molecules in solution originating from the same container and unaffected by magnetic field fluctuations may be used as a valuable tool for benchmarking quantum chemistry calculations.

We then study the effect of deuterium by detecting *J*-spectra of $^{15}\mathrm{ND}_x\mathrm{H}_{4-x}^+$ (where $x$ = 0, 1, 2, or 3). Mixtures of $^{15}\mathrm{ND}_x\mathrm{H}_{4-x}^+$ were prepared by dissolving $^{15}\mathrm{NH_4Cl}$ in solutions with varying $D_2O/H_2O$ ratios and relying on rapid equilibration of isotopologues' concentrations due to chemical exchange between $H^+/D^+$ atoms.[9]

As part of the study of $^{15}ND_xH_{4-x}^+$ ions using ZULF NMR *J*-spectra, we show that due to the hierarchy of nuclear spin-spin interactions (i.e., $\left|J_{^{15}\mathrm{NH}}\right| > \left|J_{^{15}\mathrm{ND}}\right| > |J_{\mathrm{DH}}|$), different spectral lines in the *J*-spectra have different pulse-length dependences as demonstrated in experiment, rationalized by using perturbation theory, and confirmed by numerical simulations. The value of $-2.6(1)$ Hz for the D-$^1$H scalar coupling in $^{15}NDH_3^+$ is extracted from the experiment in which various spectral lines in *J*-spectrum respond differently to magnetic field excitation pulses. We additionally discuss the possibility of creating and storing nuclear spin order in states with extended lifetimes which in principle should be achievable in the symmetric molecular cations presented in this work.

Based on results of this work, other symmetric molecular cations containing quadrupolar nuclei can be expected to produce well-resolved ZULF NMR *J*-spectra, with applications ranging from biomedicine to probing materials relevant to energy storage.

## 2. Results and Discussion

**Zero-field spectra of $^{14}$N- and $^{15}$N-ammonium isotopologues.** To understand general features of *J*-spectra at zero field, we start from the analysis of the ZULF NMR spectrum of [$^{15}$N]-ammonium, a molecular ion that does not contain quadrupolar nuclei. A *J*-spectrum of [$^{15}$N]-ammonium ion consists of two lines, one at 110.114(9) Hz and another at 183.554(7) Hz provided by the heteronuclear spin-spin coupling, $J_{^{15}\mathrm{NH}}$, between $^{15}$N and $^{1}$H spins (**Figure 1a**). Particularly, the line at 110.114(9) Hz corresponds to $(3/2)\left|J_{^{15}\mathrm{NH}}\right|$, where $J_{^{15}\mathrm{NH}}$ = -73.42 Hz is the *J*-coupling between $^{15}$N and $^{1}$H nuclei and it arises from the transitions between nuclear-spin energy levels with the total proton spin $K_{\mathrm{A}}$ = 1 (**Figure 1b**). The line at 183.554(7) Hz corresponds to $(5/2)\left|J_{^{15}\mathrm{NH}}\right|$ and arises from the transitions between states with a total proton spin $K_{\mathrm{A}}$ = 2. A corresponding high-field $^{15}$N NMR spectrum of the same sample would consist of five lines with intensities in the ratio of 1:4:6:4:1 separated by $J_{^{15}\mathrm{NH}}$ as expected from the binomial distribution for the projections of the total proton spin on the magnetic field axis.

In contrast, ZULF NMR spectrum of [$^{14}$N]-ammonium ion exhibits three well-resolved lines at 52.4021(15) Hz, 104.7741(10) Hz, and 157.171(8) Hz (**Figure 1a**). To explain this observation, one needs to consider five transitions, one in the $K_{\mathrm{A}}$ = 0 manifold, two in the $K_{\mathrm{A}}$ = 1 manifold, and two in the $K_{\mathrm{A}}$ = 2 manifold (**Figure 1c**). Two pairs of transitions overlap resulting in three distinct peaks at $J_{^{14}\mathrm{NH}}$, $2J_{^{14}\mathrm{NH}}$ and $3J_{^{14}\mathrm{NH}}$ in the ZULF NMR spectrum of $^{14}\mathrm{NH}_4^+$ ($J_{^{14}\mathrm{NH}}$ = 52.40 Hz). Since $^{15}$N and $^{14}$N have different signs of gyromagnetic ratios, $^{15}\mathrm{NH}_4^+$ and $^{14}\mathrm{NH}_4^+$ have different signs of heteronuclear $J_{\mathrm{NH}}$-couplings as indicated by inverted structures of their nuclear spin energy levels (**Figure 1b,c**).

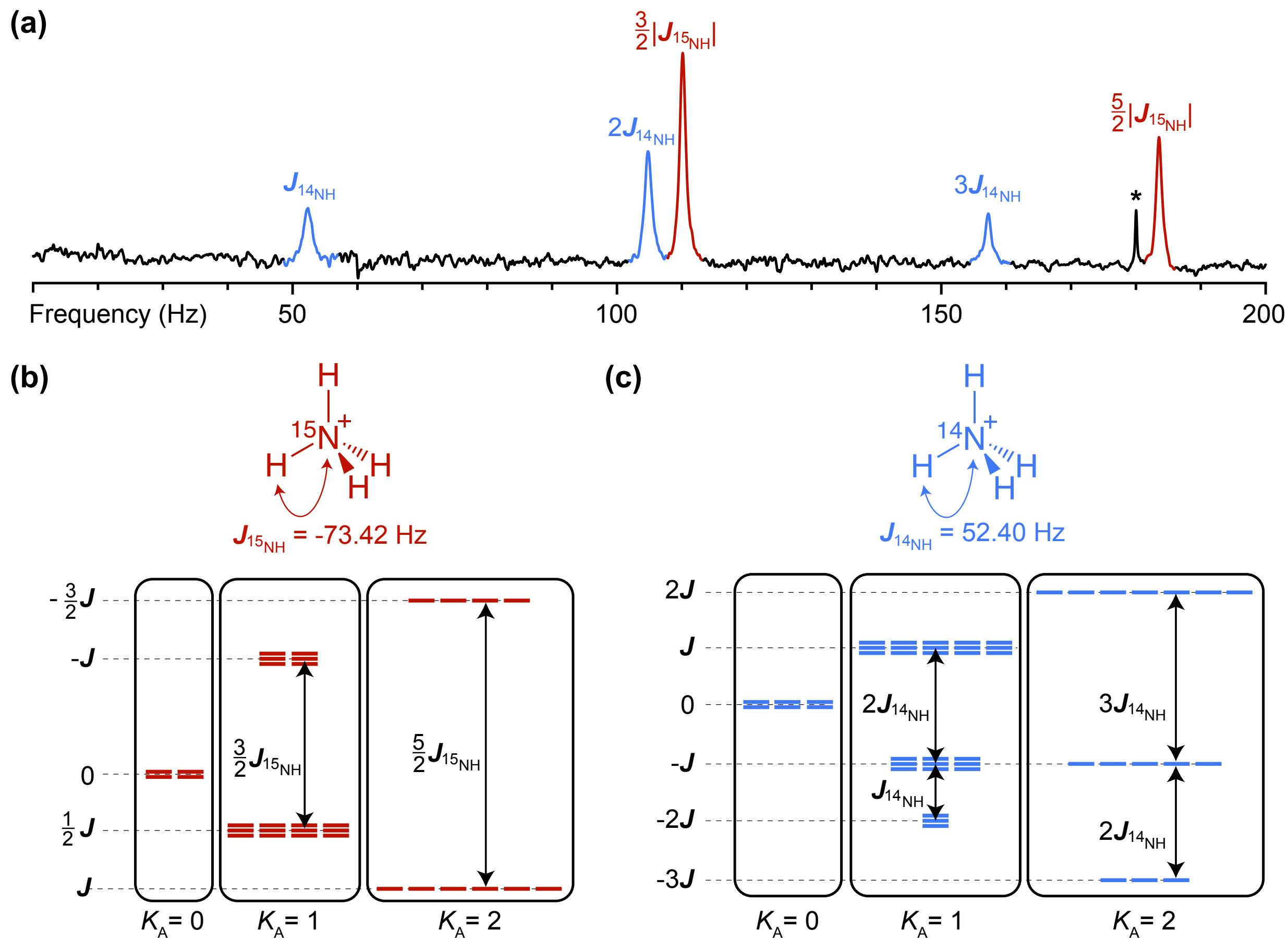

**Figure 1.** (a) Zero-field NMR *J*-spectra of the mixture of [$^{15}$N]- and [$^{14}$N]-ammonium cations ($^{15}NH_4^+$ and $^{14}NH_4^+$) in aqueous solution of $H_2SO_4$ (2 M). The asterisk depicts the third harmonic of the 60 Hz power-line frequency. (b-c) Molecular diagram and nuclear spin energy-level structure of the [$^{15}$N]-ammonium and [$^{14}$N]-ammonium cation, respectively. Observable transitions correspond to spin flips in the manifolds conserving the total proton spin $K_A = 0$, 1, and 2. The main interaction in the system is heteronuclear $J_{NH}$-coupling (in the simulation, we used $J_{HH}$=0 since it does not affect transition frequencies and, thus, do not manifest in spectra).

The mere fact of observing zero-field NMR *J*-spectrum of [$^{14}$N]-ammonium may seem surprising since $^{14}$N is quadrupolar. Indeed, a typical nuclear spin Hamiltonian of a molecule containing quadrupolar nuclei is expected to have a large contribution arising from the interaction of the nuclear quadrupole moment with the electric field gradient created by the electrons at the nucleus. For this reason, quadrupolar nuclei typically exhibit fast relaxation on timescales on the order of a few milliseconds leading to their "self-decoupling". However, in the case of $^{14}NH_4^+$ cations, the $^{14}N$ is not self-decoupled because of the high symmetry of the molecule. In general, when quadrupolar nuclei are positioned at the center of a tetrahedral or an octahedral molecule, the quadrupolar interaction is effectively switched off since electric field gradients are absent in such symmetric environments. For the same reason, the quadrupolar tensor is typically averaged out in isotropic liquids by fast molecular motion.[18] Therefore, $^{14}NH_4^+$ and, as it will be seen below, $^{15}ND_xH_{4-x}^+$, are unique spin systems containing nuclei coupled to spins with $I$ = 1 which have relaxation times on the order of seconds. However, it is important to note that in general, quadrupolar nuclei can still affect relaxation of the coupled nuclear spins and therefore can lead to broadening of spectral lines in NMR spectra.[18-19] It is of little importance at high-field NMR as measurements of ammonium cations performed at 1 tesla yielded very similar proton $T_1$ values: $T_1(^{14}NH_4^+)$ = 3.22(2) s compared to $T_1(^{15}NH_4^+)$ = 3.25(2) s. An $^{15}N$ $T_1$ measured in the same field yielded 49(3) s, see Supporting Information (SI) for details.

**Precision Measurement of *J*-couplings in $^{14}$N- and $^{15}$N-ammonium.** Indirect spin-spin coupling values between nuclei are mainly determined by the Fermi-contact interaction (i.e., probability of finding an electron inside a magnetic nucleus). Therefore, it is expected that the ratio of the experimentally measured $^{15}N$−H and $^{14}N$−H *J*-coupling values should be close to the ratio of the corresponding gyromagnetic ratios. A careful analysis revealed that the ratio of $J_{^{15}NH}$ and $J_{^{14}NH}$ coupling values differs from the ratio of the $^{15}N/^{14}N$ gyromagnetic ratios, and the difference is statistically significant. Specifically, from 36000 measurements of the $NH_4Cl$ sample containing 50:50 mixture of $^{15}N$ and $^{14}N$ isotopes, the ratio of the *J*-couplings ($\left|J_{^{15}NH}/J_{^{14}NH}\right|$) was determined to be within the range of $1.4009 - 1.4013$ (see **Table 1**) as compared to table data for gyromagnetic ratios $\left|\gamma_{^{15}\mathrm{N}}/\gamma_{^{14}\mathrm{N}}\right| = 1.4027548(5)$.[20] In the following discussion, we denote the three $J$-peaks of the $^{14}\mathrm{NH}_4^+$ cation as ①, ②, and ③, and two peaks of the $^{15}\mathrm{NH}_4^+$ cation as $\boxed{1}$ and $\boxed{2}$, for convenience.

**Table 1.** Ratios of $J_{\mathrm{NH}}$-couplings for $^{15}\mathrm{NH}_4^+$ and $^{14}\mathrm{NH}_4^+$ extracted from the analysis of 36000 averaged spectra.

| ZULF NMR peaks used for the estimation | Measured $\left|J_{^{15}\mathrm{NH}}/J_{^{14}\mathrm{NH}}\right|$ value |
|---|---|
| $(2/3) \cdot$ ($\boxed{1}$/①) | $1.4009(7)$ |
| $(4/3) \cdot$ ($\boxed{1}$/②) | $1.40108(18)$ |
| $2 \cdot$ ($\boxed{1}$/③) | $1.40103(14)$ |
| $(2/5) \cdot$ ($\boxed{2}$/①) | $1.4012(6)$ |
| $(4/5) \cdot$ ($\boxed{2}$/②) | $1.40134(15)$ |
| $(6/5) \cdot$ ($\boxed{2}$/③) | $1.40129(9)$ |

One can see that depending on which peaks in the spectrum are used to estimate the $\left|J_{^{15}\mathrm{NH}}/J_{^{14}\mathrm{NH}}\right|$ ratio, the obtained values are different. We note that although the peaks of $^{15}\mathrm{NH}_4^+$ and $^{14}\mathrm{NH}_4^+$ in ZULF NMR spectra are relatively broad (linewidths are ~1 Hz, determined by the proton exchange and not $T_2^*$), the precision of the center-frequency measurements is not limited by the linewidth.[21] To confirm if these error estimates indeed represent statistically significant variations in the measured values of $\left|J_{^{15}\mathrm{NH}}/J_{^{14}\mathrm{NH}}\right|$, we have run an additional analysis by constructing cumulative distribution functions (CDFs) using variable partitioning of the collected dataset (**Figure 2a**).

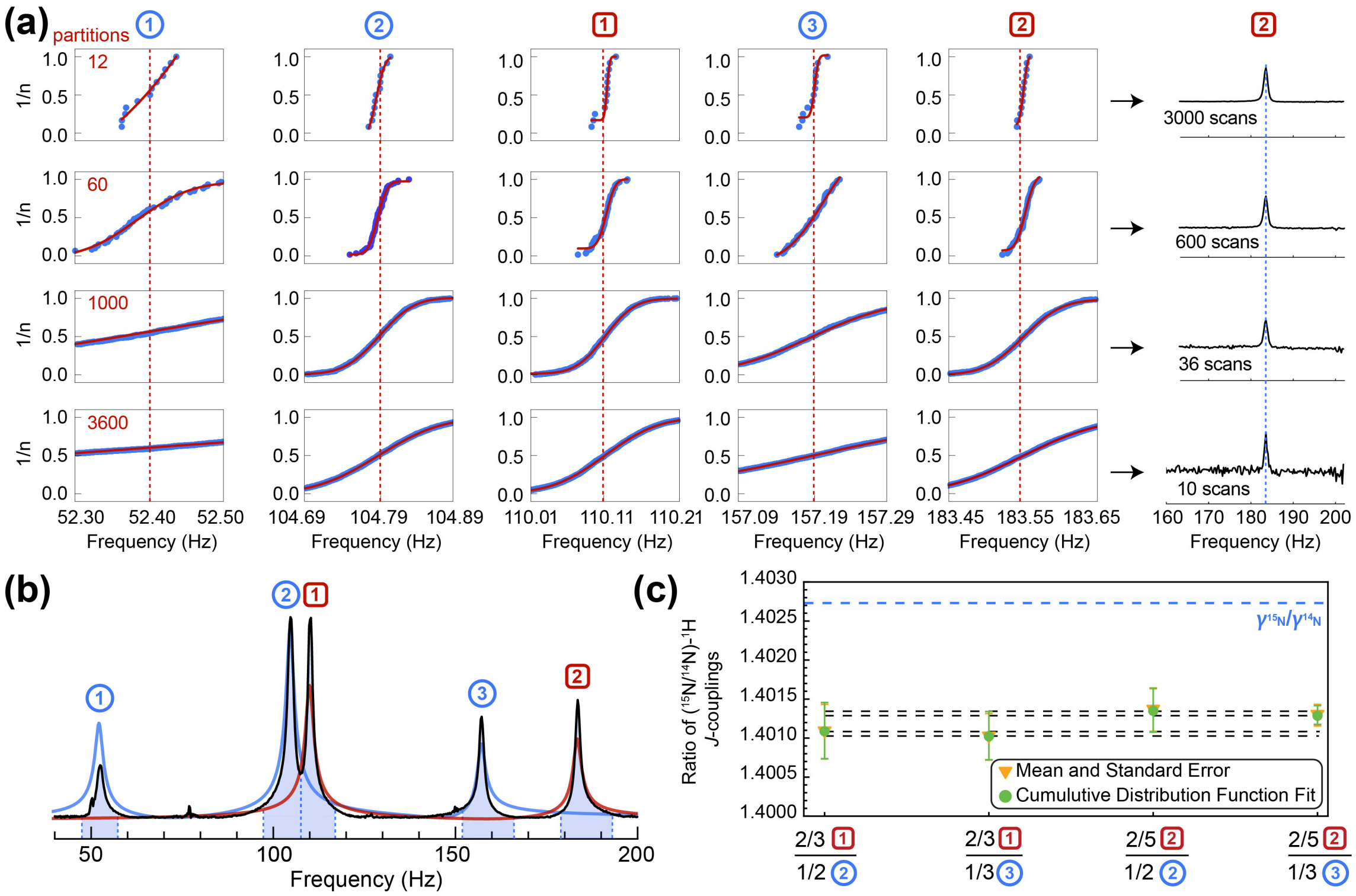

**Figure 2.** (a) Positions of the maxima extracted from peaks corresponding to isotopologues of ammonium organized in the ascending order ($1/n$ where $n$ is a partition number) generate cumulative distribution functions (CDFs); the averaged spectra with 3000, 600, 36, and 10 scans correspond to 12, 60, 1000, and 3600 partitions, respectively. (b) Zero-field NMR spectrum of [$^{14}$N]- and [$^{15}$N]-ammonium (50:50 mixture, quadrature-detected magnitude mode, average of 36000 scans) obtained with prepolarization at 2 T. (c) Values of the $J$-coupling ratios extracted from the analysis using 360 partitions (each represented by 100 averaged spectra). ZULF NMR peaks of the [$^{14}$N]- and [$^{15}$N]-isotopologues are labeled with blue circles and red squares, respectively.

Different partitions (i.e., accumulated groups of spectra) were constructed from the original data set, ranging from a single spectrum (the average of all 36000 scans, **Figure 2b**) to 3600 groups of averaged 10 scans. For each partition, the processing method described in the SI was carried out. By fitting spectral peaks using Lorentzian functions with four independent parameters (amplitude, peak center, phase, and linewidth), we extrapolated ratios of the $J$-coupling values as well as extracted the standard error of the fit; the processed dataset including the fits can be found in **Table S2**. Mean values and standard errors obtained directly from the analysis of the partitioned spectra are compared with the ones extracted from fitting the CDFs with equations for the sigmoid curve; mean values are indicated by the black dashed lines (**Figure 2c**). We observed that measured values of the peak-center frequencies shift depending on the specific data processing protocol (**Table S4**, **Figure S3**); the error bars shown in **Figure 2c** represent the corresponding systematic errors. We find that the ratio of *J*-couplings determined from the spectra is significantly different from the ratio of gyromagnetic ratios (represented by the blue-dashed line)[20] irrespective of which peaks were used for analysis. Considering average values for the extracted $J_{NH}$-couplings, these differences are an order of magnitude larger than experimental uncertainty in the measured $\left|\gamma_{^{15}N}/\gamma_{^{14}N}\right|$ value, i.e., $\Delta J = \left(\gamma_{^{14}N}/\gamma_{^{15}N}\right) J_{^{15}NH} - J_{^{14}NH} \approx -58$ mHz.[20] This is a manifestation of an isotope effect: there should be changes in the electronic structure of cations in solution that either affect Ramsey terms other than the Fermi-contact interaction,[22] or the Fermi-contact interaction itself is affected by the changes in the vibronic structure of the cations caused by $^{15}N \rightarrow {}^{14}N$ isotope substitution.[23-27] This realization highlights unexplored opportunities for the use of ZULF NMR in precision measurements of spin-spin interactions. While similar analysis could in principle be performed using conventional high-field NMR, magnetic field drifts over long experimental time window and bounds on clock precision would necessitate additional post-processing which could introduce additional systematic errors while in ZULF NMR positions of the *J*-peaks are unaffected by field fluctuations and uncertainties associated with the demodulation of the reference clock.

While some analyses report precision measurements of *J*-coupling values using conventional NMR by considering *a priori* constant distance between the peak maxima,[28] in our analysis no prior knowledge on the peak positions was assumed. Therefore, to the best of our knowledge, our study constitutes the first precision measurement of the $\left|J_{^{15}\mathrm{NH}}/J_{^{14}\mathrm{NH}}\right|$ ratio based on NMR spectroscopy unaffected by magnetic field fluctuations and additional signal post-processing to eliminate systematic errors.

Precision studies of *J*-couplings using zero-field NMR spectroscopy are proposed as a promising way to probe molecular chirality[29] and investigate parity non-conserving interactions in molecules[30]. In addition, precision measurements of molecular *J*-couplings in solution would allow comparing the results of calculating the precise molecular bond lengths by the tools of quantum chemistry with experiments, thus, making ZULF NMR a tool for benchmarking the precision of quantum chemistry calculations.[31-33] Further analysis using the modern tools of quantum chemistry could explain the exact dependence of *J*-couplings in ammonia on bond length (rotational and vibrational structure estimation). The presented measurements are relatively simple, and the investigated molecules are studied in liquid state and in the same container, thus, the effect of the dielectric properties of the environment and temperature are accounted equally for both cations.

We note that a similar analysis could also be performed with high-field NMR. However, apart from the absence of the sensitivity to inhomogeneity and magnetic-field drifts, as well as the absence of stringent requirements on the reference clock, a major additional advantage of ZULF NMR is the absence of line shifts due to chemical exchange. Indeed, upon accelerated proton exchange ZULF NMR lines of ammonium broaden without changing their center frequencies while high-field NMR peaks move toward their “center of mass”.[9, 34]

**Zero-field spectra of Deuterium Isotopologues.** To analyze the effect on *J*-spectra of quadrupolar nuclei other than $^{14}N$, we performed ZULF NMR measurements of various isotopologues of $^{15}ND_xH_{4-x}^+$ (where $x = 0, 1, 2$, or 3). While deuterium also has spin $I = 1$ (so it is quadrupolar), its quadrupole moment is significantly smaller ($0.29 \cdot 10^{-30}$ m$^2$) than that of $^{14}N$ ($1.56 \cdot 10^{-30}$ m$^2$)[18] and therefore its effect on ZULF NMR spectra is expected to be less pronounced. This expectation is now confirmed by the experiment as discussed below.

To prepare $^{15}ND_xH_{4-x}^+$ samples, the amount of deuterium was controlled by varying the ratio of $H_2O/D_2O$ in the solution. Here and below, we refer to isotopologues as *x*D where *x* represents a number of D atoms in the molecule (**Figure 3**). Since chemical exchange leads to equilibration of various ammonium-isotopologue concentrations, a molar fraction $\chi$ of the isotopologue *x*D can be derived from the fraction $p$ of the deuterium in solution as

$$\chi = \binom{4}{x} p^x (1-p)^{4-x}. \tag{1}$$

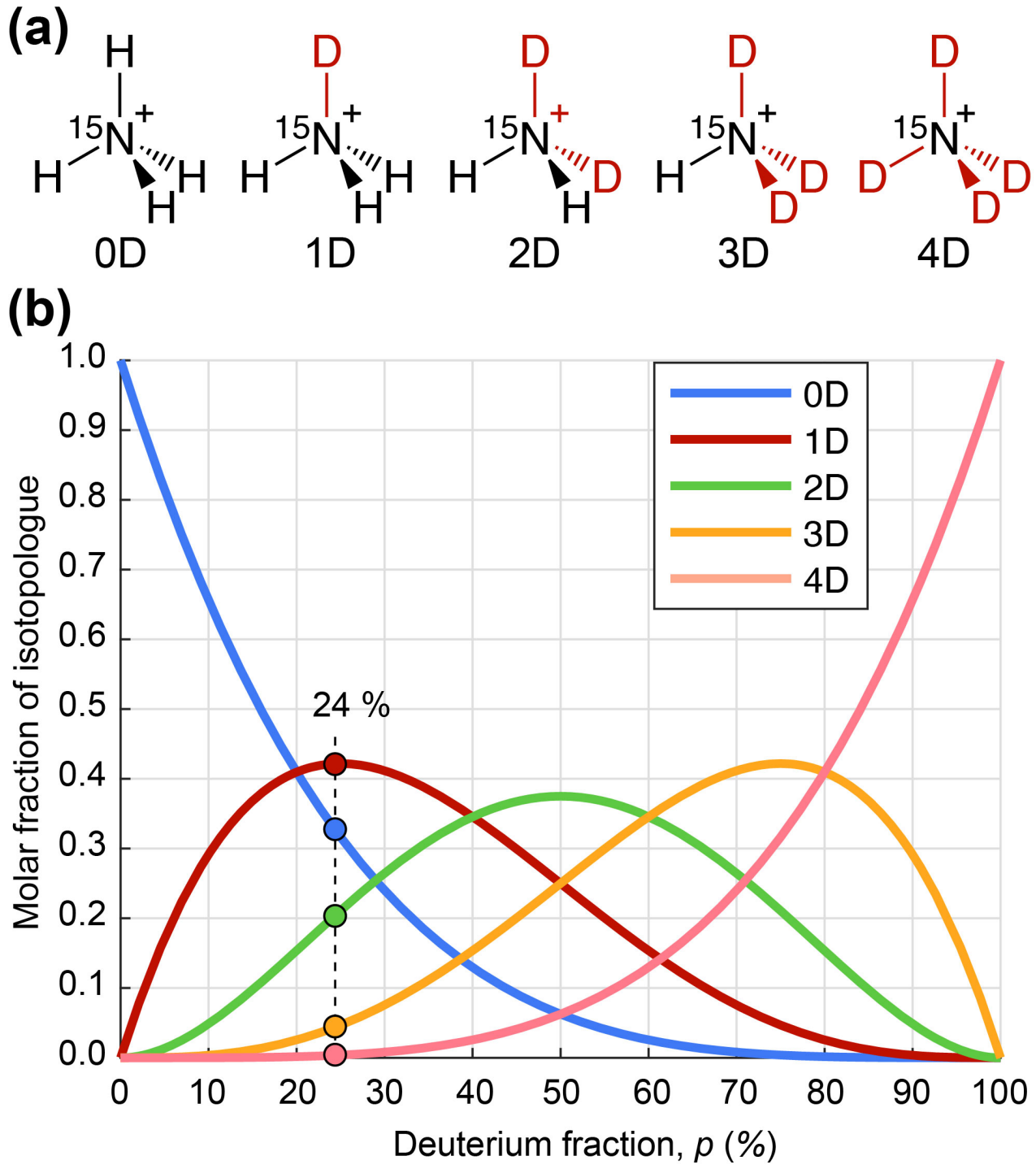

**Figure 3.** (a) Isotopologues of the ammonium cation ($^{15}ND_xH_{4-x}^{+}$) are present in the aqueous solution. The fraction of deuterium was varied by changing the amount of $H_2O$ vs. $D_2O$. Isotopologues are labeled *x*D where *x* represents a number of deuterium atoms in the molecule. (b) Equilibrium molar fractions of the isotopologues as a function of deuterium fraction in solution as determined by Eq. (1).

**Figure 4a-b** compares zero-field NMR spectra with high-field (18.8 T) $^{15}$N NMR spectra of samples containing ammonium chloride prepared by varying *p*, a fraction of deuterium in the solution. It is well known that ammonium ions undergo reversible hydrogen exchange with water.[35] High acidity of the samples can slow down the hydrogen exchange rate from ≈50,000 $s^{-1}$ down to ≈2 $s^{-1}$, thus, allowing for the molecule to exist long enough to exhibit nuclear spin coherences that can be detected with magnetometers in ZULF NMR spectroscopy.[9]

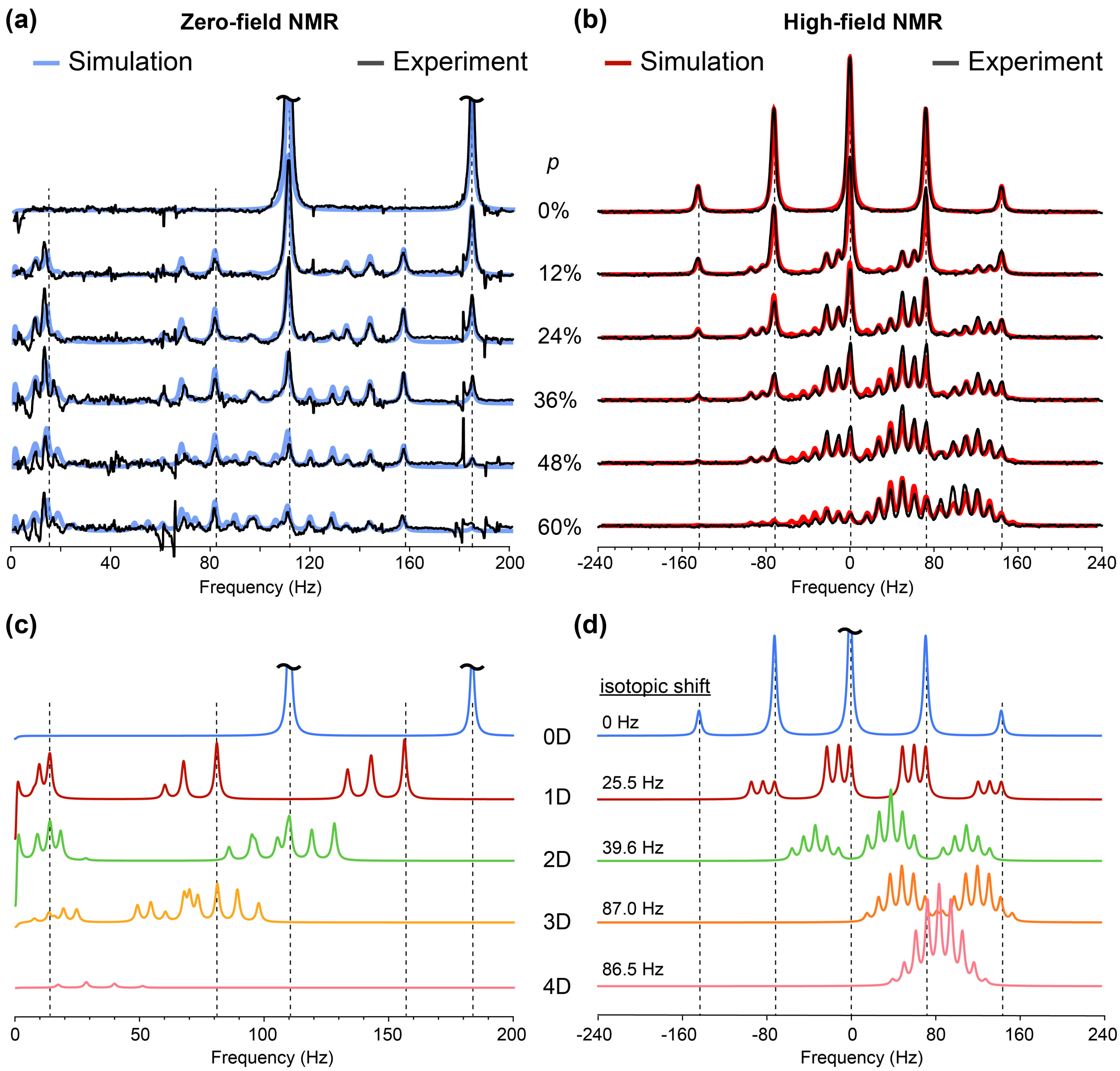

**Figure 4.** Experimentally measured (a) zero-field NMR spectra and (b) high-field (16.8 T) $^{15}$N NMR spectra of $H_2O/D_2O$ solutions containing a mixture of isotopologues of ammonium, $^{15}ND_xH^+_{4-x}$, prepared by varying the molar fraction of deuterium, $p$. Simulated NMR spectra for mixtures of the individual isotopologues calculated using weighting factors from Eq. (1) are overlaid. Simulated *J*-spectra of individual isotopologues for zero-field and high-field $^{15}$N NMR spectra are shown in (c) and (d), respectively.

Upon increase of the deuterium fraction ($p$), spectroscopic signatures of $^{15}NH_4^+$ gradually decrease; this is evident from both zero-field and high-field NMR spectra (**Figure 4a,b**). However, these spectroscopic changes are more evident in ZULF NMR spectra because the absence of chemical shift simplifies the analysis of the spectra. While each of the chemicals has its own spectroscopic signatures characterized by a unique set of frequencies in *J*-spectra (**Figure 4c**), high-field $^{15}N$ NMR spectra of isotopologues suffer from the overlap of different multiplets and changes of the $^{15}N$ chemical shift (often referred to as isotopic shift[25, 36]), due to the changes in electronic density on the nitrogen atom upon increasing a number of deuterium atoms, $x$, from 0 to 4 (**Figure 4d**).

The main reason of the decreased ZULF NMR signals for ammonium cations with larger deuterium content is smaller initial magnetization of the sample due to decreased number of protons. Indeed, in a simple two-spin case, ZULF NMR signal is proportional to the difference between gyromagnetic ratios of the *J*-coupled nuclei.[5] In case of many-spin systems composed of more than two types of heteronuclei and prepolarization at high field, the detected signal depends on a combination of gyromagnetic ratios determined by the exact nuclear spin topology of the molecule.[37] However, the general trend is that the main magnetization-contributing nuclei, protons, predominantly contribute to the observed signal as they have significantly larger gyromagnetic ratio compared to other spins in the system, for example, $\left|\gamma_{^1\mathrm{H}}/\gamma_{^{15}\mathrm{N}}\right| \approx 10$ and $\left|\gamma_{^1\mathrm{H}}/\gamma_{^2\mathrm{H}}\right| \approx 6.5$.

The fact that the signal intensity in the *J*-spectrum of the molecule depends not only on the concentration but also on the exact spin-type composition and spin topology presents a general limitation of ZULF NMR spectroscopy for chemical analysis. This contrasts with the case of conventional high-field NMR where the detected signal is proportional to molecular concentration independent of topology.[38] Therefore, for analytical purposes, ZULF NMR spectral features of different chemicals should be calibrated (and/or simulated) for direct spectroscopic comparison. The agreement between the experimentally obtained and simulated ZULF NMR spectra is excellent for the $^{15}NH_4^+$, $^{15}NDH_3^+$, and $^{15}ND_2H_2^+$ molecules (**Figure 4**). Signatures of other isotopologues $^{15}ND_xH_{4-x}^+$ (where $x$ is 3 and 4) are noticeable in the spectra as well, however, their signal is lower and, therefore, precise spectroscopic assignment is less straightforward.

We note that another possible reason of the decreased NMR signal for the molecules with larger deuterium content is scalar relaxation of the second kind (SR2K).[39] In the context of ultralow-field NMR, SR2K may lead to accelerated relaxation of overall polarization as well as coherences involving spin-½ nuclei mediated through *J*-couplings with the quadrupolar nucleus. Accelerated relaxation is especially pronounced in the regime where nuclei are strongly coupled.[40] In simple words, when spins are strongly coupled, they tend to relax together rather than separately, and the presence of a quadrupolar spin relaxing on the timescale of milliseconds acts as a relaxation sink for the rest of the spin system. For example, fast liquid-state polarization decay was observed for hyperpolarized [5-$^{13}$C]-glutamine during the transfer to an MRI scanner when the transfer field was below 800 μT and was attributed to the SR2K caused by the fast-relaxing quadrupolar $^{14}$N-nucleus adjacent to the $^{13}$C nucleus in the amide group.[41] Another study showed that partially deuterated ethanols have significantly larger linewidth in near-zero-field Larmor-precession experiments implying a strong SR2K contribution to $^{1}$H relaxation rates despite relatively weak coupling constants on the order of 1-2 Hz.[14] However, the situation is different for $^{15}ND_xH_{4-x}^+$ isotopologues investigated in our work because, first, we do observe their zero-field *J*-spectra and, second, the linewidth of resonances is determined by the intermolecular chemical exchange and not the SR2K mechanism. The absence of an overwhelming SR2K contribution may potentially be explained by the fact that the quadrupolar relaxation rates of the spin state imbalances in

$^{15}\mathrm{N\,D}_x\mathrm{H}_{4-x}^{+}$ are determined by the correlation time of the molecular rotations and thus, symmetry properties of these molecules must be accounted for. An exact explanation of the relaxation phenomena at zero field requires additional analysis which lies beyond the scope of this paper. We note that long-lived spin states in molecules containing deuterium, were demonstrated for the deuterated methyl groups.[42]

As a general rule, *J*-spectra featuring couplings to quadrupolar nuclei should be observable at ZULF conditions if the corresponding *J*-coupling is manifested in conventional high-field NMR spectra. For example, as seen from both **Figure 4a** and **4b**, a splitting of the NMR lines due to $J_{^{15}\mathrm{ND}}$ is clearly observed in the zero-field and high-field spectra.

**Evaluation of $J_{\mathrm{HD}}$ from ZULF NMR spectra.** High-field $^{1}$H NMR measurements cannot directly resolve $^{1}$H-D *J*-coupling in $^{15}\mathrm{ND}_x\mathrm{H}_{4-x}^{+}$ systems. For example, **Figure 5a** shows a $^{1}$H NMR spectrum of the [$^{15}$N]-ammonium solution with a 24% fraction of deuterium atoms. One can clearly see a series of doublets split by heteronuclear $^{1}$H-$^{15}$N *J*-coupling of -73.4 Hz shifted to lower field for each isotopologue $^{15}\mathrm{N\,D}_x\mathrm{H}_{4-x}^{+}$ as $x$ increases from 0 to 2; however, the fine structure corresponding to $^{1}$H-D coupling is not visible and only broadening of the peaks is observed. To investigate opportunities of resolving $^{1}$H-D *J*-coupling with zero-field NMR techniques, we numerically simulated zero-field spectra for deuterated ammonium isotopologues as well as analyzed their energy-level structures using the perturbation theory (see SI).

For all isotopologues $^{15}\mathrm{ND}_x\mathrm{H}_{4-x}^{+}$ (where $x = 1 - 3$) one can distinguish regions of low-frequency (0-30 Hz) peaks and high-frequency peaks (120-170 Hz). High-frequency peaks correspond to transitions within the strongly coupled subsystem consisting of $^{15}$N and $^{1}$H spins when total deuterium spin remains unchanged. Low-frequency peaks correspond to the deuterium spin flips where the strongly coupled $^{15}$N-$^{1}$H subsystem remains unperturbed. Due to the hierarchy of interactions in the molecule ($|J_{^{15}\mathrm{NH}}| > |J_{^{15}\mathrm{ND}}| > |J_{\mathrm{HD}}|$) these groups of peaks are expected to respond differently to the magnetic pulse excitation and thus, by plotting their integrals as a function of the pulse length, one can extract information about subtle spin-spin interactions which otherwise would have taken more sophisticated multinuclear high-field NMR analyses (**Figure S3**). By performing such analysis, we were able to constrain the value of $J_{\mathrm{HD}}$ to -2.6(1) Hz.

Interestingly, unlike in the case of high-field NMR, signs of *J*-couplings alone can significantly affect the positions of the ZULF NMR peaks (**Figure 5b**). Using a sample with deuterium fraction of 24% as an example, we simulated the *J*-spectra by varying signs of $J_{^{15}\mathrm{NH}}$, $J_{^{15}\mathrm{ND}}$, and $J_{\mathrm{HD}}$ (one should note that reversal of signs for all spin-spin couplings at the same time would result in the same spectral pattern, however, the value of $J_{^{15}\mathrm{NH}}$ is known to be negative, see **Figure 1**). Especially pronounced are effects of the *J*-coupling sign change on the intensity of the high-frequency peaks of $^{15}\mathrm{NDH}_3^+$ as well as on the nutation pattern for the low-frequency peaks.

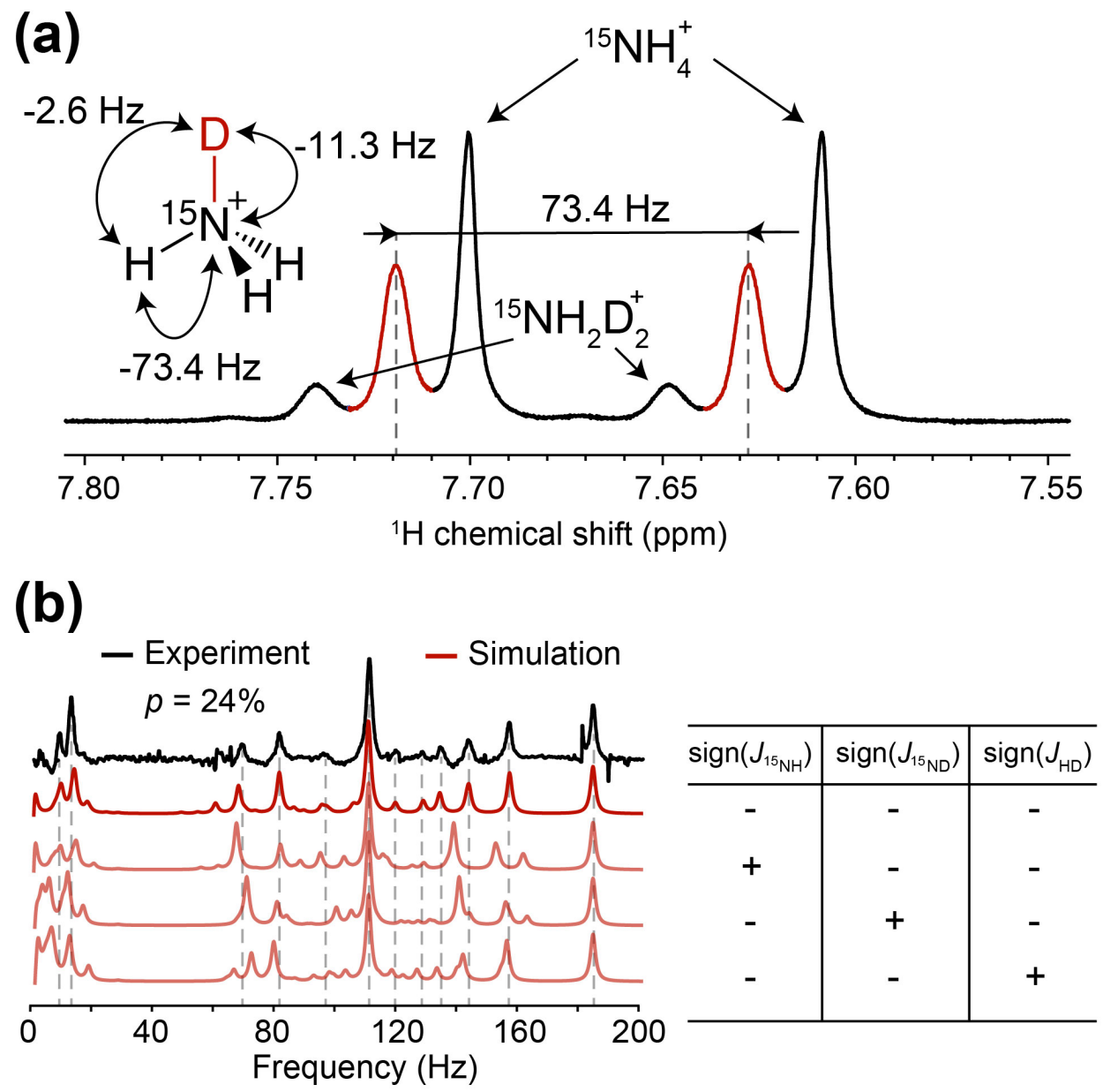

**Figure 5.** (a) $^1$H NMR spectrum (800 MHz) and (b) ZULF NMR spectrum (black) of $^{15}ND_xH^+_{4-x}$ solution with deuterium fraction $p$ = 24%. In (b), calculated zero-field $J$-spectra of a mixture of isotopologues is shown in red, corresponding signs of $J$-couplings are presented on the right. Given insensitivity to global sign changes, only the four non-redundant permutations from the total eight possible combinations of $J$-coupling signs are shown.

Numerical calculations support the experiments which demonstrate that high-frequency and low-frequency peaks have maximal intensity at different pulse lengths. This ability to maximize some parts of the $J$-spectra while suppressing others opens possibilities for on-demand spectral editing in zero-field NMR spectroscopy using duration of the excitation pulse alone. Indeed, in case when complex chemical mixtures are studied using ZULF NMR tools, this feature could yield an additional degree of freedom for disentangling the spectral complexity. In principle, the experiment can be performed in a two-dimensional (2D) manner where Fourier transformation in the indirect (pulse length) dimension would give separate peaks in the 2D spectrum revealing additional information about spin topology that may simplify molecular characterization. We note that experiments with $^{14}ND_xH_{4-x}^+$ could also be carried out, additionally avoiding isotopic enrichment with $^{15}N$ nuclei. However, since their ZULF NMR spectra would have additional splittings due to the presence of spin-1 $^{14}N$ nucleus, SNR would be further reduced making the analysis more time consuming.

## 3. Outlook

Symmetric ions such as $^{14}NH_4^+$ and $^{15}NH_4^+$ represent an interesting class of molecules for storage of the nuclear spin order. Indeed, less symmetric analogs of ammonium, methyl groups, are known to support long-lived spin states (LLSS)[43-44] and long-lived coherences (LLC)[45] due to the fact that transitions between the states of different irreducible representations are symmetry-forbidden for intramolecular dipole-dipole relaxation.[42] While LLSS have already been observed in $^{13}CH_3$/$^{13}CD_3$ groups,[42, 46-47] the search of spin orders with extended lifetimes in $^{14}NH_4^+$/$^{15}NH_4^+$ is certainly warranted. The ability to store nuclear spin coherences for a long time will enable enhanced spectral resolution (i.e., narrow lines) in zero-field NMR. In addition, the existence of the LLSS in symmetric ions could allow preserving hyperpolarization on the timescale significantly longer than $T_1$. One should note that hyperpolarization of ammonia and its derivatives was demonstrated by using signal amplification by reversible exchange (SABRE) approach.[48-49]

Other molecules containing quadrupolar nuclei in tetrahedral or octahedral environments may serve as subjects for future investigation by the ZULF NMR spectroscopy. For example, $BF_4^-$ and $PF_6^-$ are counter-ions in various organometallic complexes and ionic liquids, as well as electrolytes in batteries or electrochemical double-layer capacitors.[50] Thus, zero-field *J*-spectroscopy of symmetric ions including ones with quadrupolar nuclei can find potential application for probing Bronsted acid sites in zeolites, lipid bilayers, and the hydrated ionomer membranes used in fuel cells.[51]

Some molecules containing quadrupolar nuclei may find applications in biomedicine. For example, choline and its derivates are proposed as biomarkers for cancer diagnosis and response to treatment using hyperpolarized *in vivo* NMR spectroscopy.[52-53] While most of the research to date is focused on polarizing $^{15}$N-labeled choline derivatives, based on the findings of this study, preparation of $^{14}$N hyperpolarization in non-labeled (i.e., $^{14}$N-containing) molecules should be feasible as well. While $T_1$ of $^{14}$N sites in choline is still relatively short (~4 s),[54] which may limit applicability of the conventional approach, the long-lived spin states can provide a platform for the storage of enhanced population imbalance. At the same time, ZULF NMR detection of the characteristic spectroscopic signatures does not require hyperpolarization and therefore, small molecules containing $^{14}$N in semi-symmetric environments can be used as tracers for *in vivo* ZULF NMR studies. For example, acetylcholine is an important neurotransmitter and monitoring its concentration by optical magnetometers could yield valuable metabolic information.[55] Feasibility for the realization of these ideas is supported by the fact that $^{1}$H–$^{14}$N heteronuclear correlation (HSQC) spectroscopy of choline-containing compounds in solutions has been successfully demonstrated *in vitro*[54] and *in vivo* (*i.e.*, "rule of thumb" for quadrupolar nuclei is fulfilled).[56-57]

## 4. Conclusions

To summarize, we report zero-field NMR measurements of molecules featuring the *J*-coupling to quadrupolar nuclei. Solutions containing different isotopologues of ammonium cations, $^{14}\mathrm{NH}_4^+$ and $^{15}\mathrm{ND}_x\mathrm{H}_{4-x}^+$ (where $x$ = 0, 1, 2, or 3) were studied and their zero-field NMR *J*-spectra were measured. Molecules containing a larger number of deuterons compared to protons are characterized by a lower intensity of resonances in *J*-spectra as attributed to the decreased number of protons in the molecule (less overall magnetization) and not the scalar relaxation of the second kind. For the spin systems containing more than two types of magnetic nuclei, different groups of peaks in *J*-spectra have a different dependence of the magnetic pulse length provided a suitable hierarchy of nuclear spin-spin interactions ( $\left|J_{^{15}\mathrm{NH}}\right| > \left|J_{^{15}\mathrm{ND}}\right| > |J_{\mathrm{HD}}|$ ). Spin-spin coupling values and their signs are extracted for $^{15}\mathrm{NDH}_3^+$ and $J_{\mathrm{HD}}$ value was determined to be $-2.6(1)$ Hz. First to date precision measurement of $\left|J_{^{15}\mathrm{NH}}/J_{^{14}\mathrm{NH}}\right|$ is performed and the value differs within a range of $1.4009 - 1.4013$. The statistically significant difference of this value with the measured ratio of $^{15}\mathrm{N}/^{14}\mathrm{N}$ gyromagnetic ratios is reported, this primary isotope effect is attributed to the changes in electronic structure and rovibronic energy potential for the cations. Such subtle differences in electronic structure for $^{15}\mathrm{NH}_4^+$ and $^{14}\mathrm{NH}_4^+$ are manifested via *J*-spectra and for the first time are reported for the molecules having identical environment in the liquid phase. Detection of ammonium ions by atomic magnetometers at zero field provides specificity toward various isotopologues and paves the way for studying other symmetric molecules containing quadrupolar nuclei with potential applications in biomedicine and energy storage.

## 5. Methods

Zero-field *J*-spectra of the ammonium samples (**Figures 1**, **4-5**, and **S3**) were recorded using a home-built ZULF NMR spectrometer (incorporating $^{87}$Rb atomic magnetometer inside multilayer magnetic shielding). Physical principles of operation, construction, and calibration of the instrument are described in detail in Ref. 6. Each *J*-spectrum is a result of 100 averages, polarization time is 30 s in the field of 2 T, shuttling time to the zero-field chamber is ~0.5 s. During the shuttling, no guiding field was applied. To generate an observable ZULF NMR signal, a magnetic pulse of variable length was applied in a direction of the magnetometer sensitive axis.

The data presented in **Figure 2** was recorded using a home-built ZULF NMR spectrometer (with multilayer magnetic shielding to attenuate Earth's magnetic field $10^6$-fold based on commercial sensors[58-59] (QuSpin QZFM Gen-2; 4×4×4 mm$^3$ Rb vapor cell) with a gradiometric configuration.[60] A total of 36000 scans were acquired, with 6 s polarization time at 1.8 T. The shuttling time to the zero-field chamber is ~0.5 s. The measurement protocol consisted of the following steps: (i) a piercing solenoid was used to apply 100 μT field for 0.4 s during the sample shuttling (using pneumatic setup); (ii) during the same period, an additional 100 μT in the shuttling direction was also applied using a Helmholtz-coil pair; (iii) during 0.1 s, the piercing solenoid was switched off while the Helmholtz field remained (the previous two steps account for the shuttling time of 0.5 s); (iv) to ensure adiabaticity towards the sensitive axis (orthogonal to the shuttling direction) we performed a field "crossing", where the field of 100 μT was ramped down to zero within 50 ms while simultaneously a different Helmholtz coil along the magnetometer sensitive axis ramped the field from zero to 100 μT; (iv) finally, the field was non-adiabatically ($<$10 μs) switched off, resulting in an observable signal.

Solutions of $^{15}NH_4Cl$ (Sigma-Aldrich 299251) and/or $^{14}NH_4Cl$ (Sigma-Aldrich 254134) were prepared in a 6 M concentration by dissolving them in de-ionized water followed by the addition of 98% (mass percent) sulfuric acid (Sigma-Aldrich) to a final concentration of 1.9 M. Solutions were then purged of dissolved oxygen by bubbling with nitrogen gas for 10 minutes at 1 atm. The samples were then flame sealed in 5 mm NMR tubes under vacuum. Sample volumes were 300 μL for ZULF measurements and 500 μL for high-field measurements.

Solutions of $^{15}ND_xH_{4-x}^+$ were prepared by dissolving $^{15}NH_4Cl$ (320 mg per vial) into a mixture of distilled water and $D_2O$ (Cambridge Isotope Laboratories DLM-4-99.8-1000) according to a desired deuterium fraction ($p$) followed by the addition of concentrated sulfuric acid (98% by mass) to yield a total volume of 1.0 mL with 6 M $^{15}NH_4Cl$ and 1.9 M $H_2SO_4$. 300 μL portions of solution were transferred to standard 5 mm NMR tubes and subsequently sealed with a flame.

Zero-field spectra of $^{14}NH_4^+$ and $^{15}ND_xH_{4-x}^+$ were calculated using a code written in Wolfram Mathematica (see supplementary file "ZULF_NMR_simulations.nb"). The initial state of the density operator corresponded to spins being polarized along the $z$-axis according to Boltzmann distribution considering the gyromagnetic ratio of each nucleus. Liouville-Von-Neumann equation was solved numerically to calculate the evolution of the density matrix under the action of zero-field Hamiltonian. The observable operator corresponded to the total magnetization produced by all the spins along the $z$-axis. The time-dependent density matrix was projected onto the observable operator to provide the time domain signal, which was then Fourier transformed to produce a zero-field spectrum for a given isotopomer. Additionally, an analytical perturbation model was used (see SI) to calculate eigenfrequencies and eigenstates of the zero-field Hamiltonian for each of the isotopomers of ammonia cation and to cross-check numerical calculation and to gain a better understanding of underlying physics in the studied molecular systems.

**Data availability**

Fitting data is completely defined by the parameters reported in Tables S1-S2.

**Code availability**

Simulations of ZULF NMR spectra are provided in a separate supplementary file "ZULF_NMR_simulations.nb".

**Acknowledgments**

We acknowledge the financial support by the Alexander von Humboldt Foundation in the framework of the Sofja Kovalevskaja Award. DB acknowledges support by the Cluster of Excellence Precision Physics, Fundamental Interactions, and Structure of Matter (PRISMA+ EXC 2118/1) funded by the DFG within the German Excellence Strategy (Project ID 39083149).

**Contributions**

D.A.B. conceptualized research and wrote the first draft of the manuscript. R.P.F., K.F.S., E.V.D. updated the main manuscript and the Supplementary Material. R.P.F., J.W.B., M.R., E.V.D., K.F.S, D.A.B. collected data. R.P.F., K.F.S., J.W.B., T.S., A.P., D.B., D.A.B. contributed to the analysis and interpretation. D.B. and D.A.B. supervised the research. All authors edited and approved the final manuscript.

**Ethics declarations**

The authors declare no competing interests.

# *Supporting Information:*
# *Zero-Field J-spectroscopy of Quadrupolar Nuclei*

Román Picazo-Frutos,[a] Kirill F. Sheberstov,[a,b] John W. Blanchard,[a,c] Erik Van Dyke,[a]

Moritz Reh,[d,e] Tobias Sjoelander,[f,g] Alexander Pines,[g]

Dmitry Budker,[a,d] and Danila A. Barskiy*[,a,g]

[a]Johannes Gutenberg-Universitat Mainz, 55128 Mainz, Germany

Helmholtz-Institut Mainz, GSI Helmholtzzentrum, 55128 Mainz, Germany

[b]Department of Chemistry, École Normale Supérieure, PSL University, Paris, France

[c]Quantum Technology Center, University of Maryland, College Park, MD, USA

[d]Department of Physics, University of California – Berkeley, Berkeley, California 94720, USA

[e]Kirchhoff-Institut für Physik, Universität Heidelberg, 69120 Heidelberg, Germany

[f]Department of Physics, University of Basel, Klingelbergstrasse 82, Basel, CH-4056, Switzerland

[g]Department of Chemistry, University of California - Berkeley, California 94720-3220, USA

Materials Science Division, Lawrence Berkeley National Laboratory, Berkeley, California 94720-3220, USA

## A. Analysis procedure for extracting *J*-coupling ratio

All of the scans of the [$^{15}$N]- and [$^{14}$N] ammonium mixture (**Figure 1**) were subject to the same data analysis. Two commercially available magnetometers (QuSpin QZFM Gen-2; 4×4×4 mm$^3$, Rb vapor cell) were used to detect the signal in a gradiometer configuration. The difference between the two channels of the sensitive axis was taken as gradiometric output signal, thus, enhancing the signal (by a factor ≈1.5 determined by the geometry of the sample and the exact location of the sensors) while simultaneously removing any common-mode noise. The experimental setup details can be found elsewhere.[1]

The data were acquired at a sampling frequency of 50 kHz and later downsized to 1 kHz (to follow the Nyquist condition to measure frequencies of up to 500 Hz) by an averaging-down procedure which removed high-frequency noise.[1] The NMR spectra were obtained from the time-domain signals via the usual Fourier transformation used in conventional high-field NMR. An extra step is added by scaling up by a factor of two every point of the time-domain signal except the initial point to get rid of baseline-offset artifacts, described recently in Ref. [2]. The data were zero-filled by the same amount of data points $t_{\mathrm{acq}}$ (zero-filling factor = 2) and no line broadening was added.

The resulting spectra exhibit a non-flat baseline attributed to a significant DC-offset component (observed at 0 Hz and not fully removed by the gradiometric detection, see below) and low-frequency fluctuations in the time-domain signal stemming from the temperature-lock PID readjustment of the optically pumped magnetometer (OPM) following the excitation pulse. This baseline profile poses challenges when fitting *J*-coupling frequencies. To address this issue, several steps have been implemented; here, we outline all of them:

(i) Subtraction of the two gradiometer channels of the two OPMs to enhance the signal while reducing common-mode noise.

(ii) Application of a moving average (over 25 subsequent points) subtraction to eliminate low-frequency oscillations, ensuring that high frequencies (>50 Hz) remain unaffected.

(iii) Removal of the initial 50 ms of data (since the OPM requires this time to regain sensitivity after saturation caused by the magnetic-field excitation pulse) leaves us with useful data points corresponding to $t_{\mathrm{acq}} = 2.048$ s (the sampling rate is 1 kHz).

(iv) Subtraction of analytical functions and sinusoidal components at 50 Hz and overtones from the original data in the time domain. Specifically, we employed the following fit model: $model = \{"exponential\ decays"\} + \{"power - line\ noise\ sinusoidals"\} + \{"decaying\ sinusoidals"\} + \{"3rd - degree\ polynomial"\}$, resulting in a fit with a coefficient of determination $R^2 = 0.991$.

(v) Application of a 3$^{\mathrm{rd}}$-order bandpass filter to remove low-frequency noise, with low and high cutoff frequencies set at 20 Hz and 1000 Hz, respectively.

(vi) Additional baseline removal in the frequency domain by interpolating the baseline outside the regions of interest and subtracting it from the Fourier data. The complex data were phase-corrected to compensate for the initial dropped points and experimental imperfections.

(vii) Conversion of the spectra into magnetic-field units following the OPM calibration conversion factor of $0.9\ \mathrm{V/nT}$.

These steps applied to the entire dataset of 36000 spectra ensure the accuracy and quality of the analysis. In steps (ii) and (v), the applied filters had a low-frequency cutoff significantly lower than the lowest-frequency NMR peak. The baseline-reduction steps in frequency domain (step vi) included frequency ranges far from Lorentzian peaks.

It is important to note that the standard phase correction typically applied to high-field NMR spectra (0th and 1st order) proved insufficient for achieving universally positive absorptive lines in the ZULF NMR spectra. To address this, we implemented an interpolation of a frequency-dependent phase function $\varphi(\nu)$ enabling an absorptive phase across all relevant NMR peaks. The reason of this phase accumulation is currently under investigation. This modeling approach ensured the absence of baseline distortions introduced by high-order phase corrections.[3]

To obtain enough statistical points for extracting *J*-coupling frequencies, we performed identical shuttling experiments with prepolarization at 2 T for $N_0 = 36000$ scans. Different partitions were constructed from the original dataset, ranging from a single scan (the average of all 36000 scans) to groups of 10-averaged scans each (3600 partitions). For each partition, the whole processing procedure described above was carried out, with the same values of phase correction and baseline interpolation of the average of all scans were used for all of the partitions. We then fit the real part of the sum of five complex Lorentzians (each with variable phase parameter), to our complex data. From this fitting, we extrapolate the *J*-coupling values as well as the standard error of the fit.

Once we have extracted five *J*-coupling frequencies from the fit for each partition (**Figure 2**), we constructed the cumulative distribution function (CDF)[5] by (i) sorting all the values $\{\nu\}$ in ascending order in the horizontal axis and (ii) constructing $1/n$ points in the vertical axis, where $n$ is the number of partitions (e.g. for groups of 3000 averaged scans, the partition number is $n = N_0/3000 = 12$, thus, the vertical axis ranges as $\{1/12, 2/12, \ldots, 11/12, 1\}$ ). These so-called sigmoid curves are fit with a CDF of a normal distribution with mean $\mu$ and standard deviation $\sigma$:

$$\frac{1}{2} k_1 \mathrm{Erfc}\left(\frac{\mu - \nu}{\sqrt{2}\sigma}\right) + k_2 \,, \tag{S1}$$

where $k_1$and $k_2$ are constant and Erfc is the error function. The result can be found in **Figure S1**. Note that only after the partition with more than four points (fifth row and onwards), the fit is adequate. One can note that the sigmoids broaden with an increasing number of partitions since the fit of each individual averaged scan becomes less accurate (higher-order partitions have lower signal-to-noise ratio).[6]

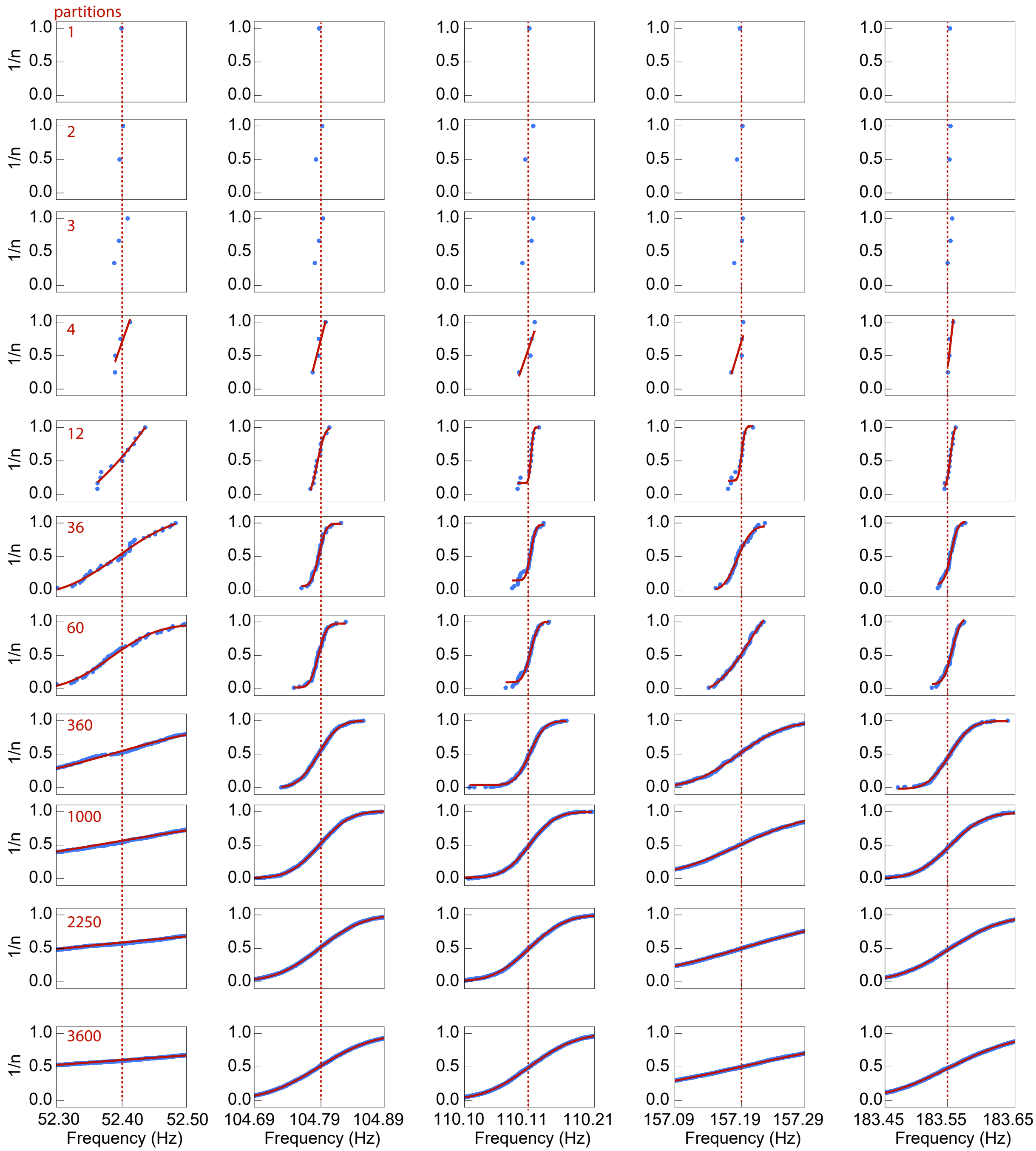

**Figure S1.** Sigmoid curves including fit with an increasing number of partitions.

From the sigmoid fits (**Figure S1**), we can extract the mean value of each frequency and the standard error of the mean (standard deviation divided by the square root of the number of measurements). Since there are several peaks corresponding to [$^{14}$N]- and [$^{15}$N]-ammonia, there are 6 different ways to extract the ratio of *J*-couplings. The results can be found in **Figure S2**.

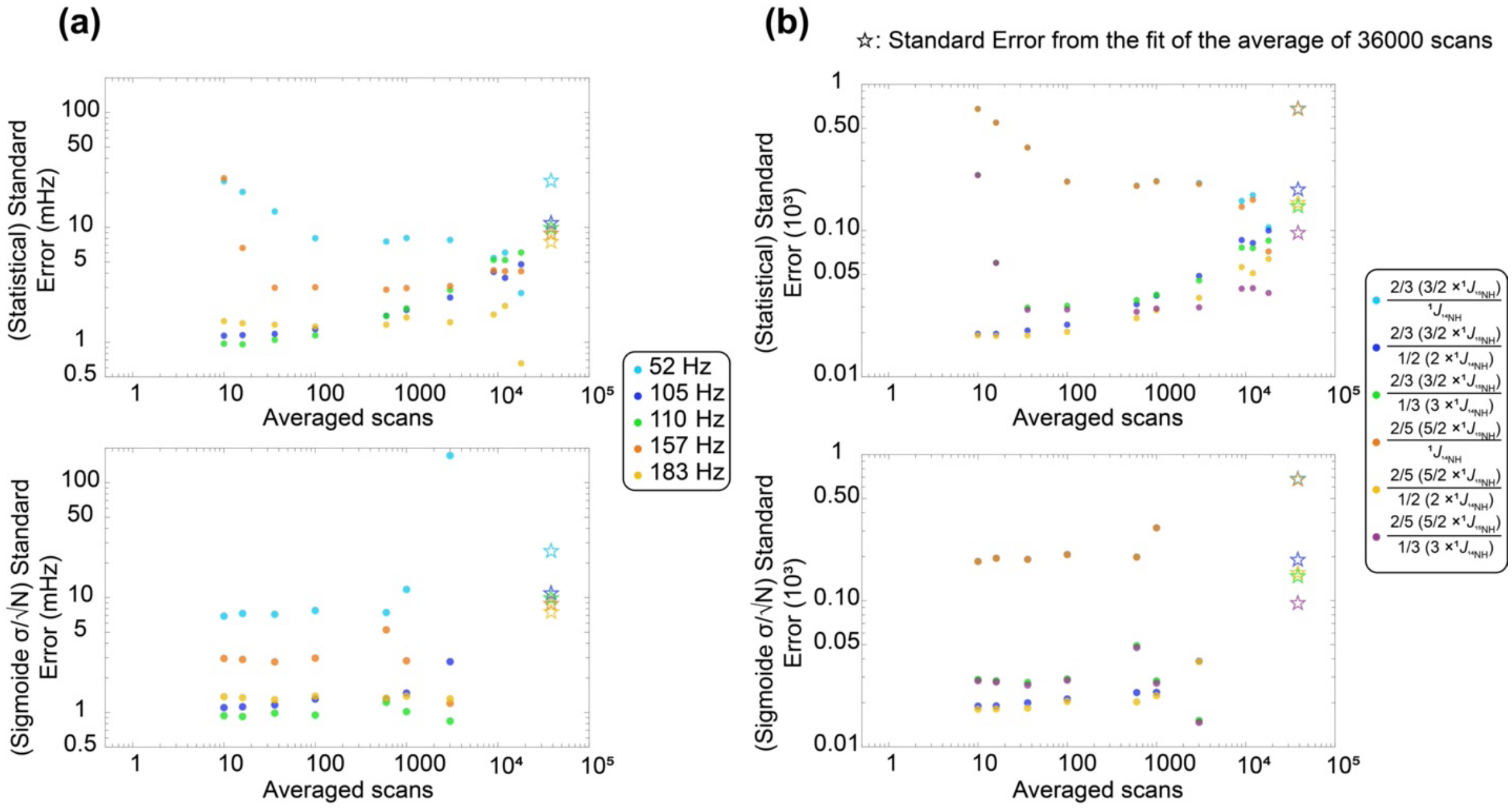

**Figure S2.** Standard error of the **(a)** *J*-coupling frequencies and **(b)** ratio of *J*-coupling frequencies extracted for different partitions of averaged scans. The top row shows values of the standard error determined from individual measurements (statistics) and the bottom shows the value of the standard error extracted from the fits of sigmoid curves.

Notice how it is possible to extract frequency values and statistical parameters for groups of 2, 3, and 4 averaged scans in the top three rows of **Figure S1**, however, the sigmoid fit is only possible from the next partition (12 groups of 3000 averaged scans), since one needs more points to properly fit Eq. S1 to the data. The value obtained from the direct fit of all averaged scans is added with a ☆ symbol for comparison.

The standard error of the ratio of *J*-couplings in **Figure S2b** was obtained from the values in **Figure S2a** using error propagation formula:

$$\sigma_{\frac{J_{15_{\mathrm{NH}}}}{J_{14_{\mathrm{NH}}}}} \approx \left(\frac{J_{15_{\mathrm{NH}}}}{J_{14_{\mathrm{NH}}}}\right) \sqrt{\left(\frac{\sigma_{J_{15_{\mathrm{NH}}}}}{J_{15_{\mathrm{NH}}}}\right)^2 + \left(\frac{\sigma_{J_{14_{\mathrm{NH}}}}}{J_{14_{\mathrm{NH}}}}\right)^2} \ . \tag{S2}$$

The numerical values of **Figure S2** are also tabulated in **Table S2** for completeness. The *J*-couplings and their ratios are expressed in the order introduced in the **Table S1**.

**Table S1**. Frequencies of the zero-field peaks and their relationship to *J*-coupling values. Peaks of $^{14}NH_4^+$ and $^{15}NH_4^+$ are denoted as ①, ②, ③, and $\boxed{1}$, $\boxed{2}$, respectively.

| Peak | Corresponding frequencies (Hz) | Relationship to *J*-coupling | Ways of extracting $\left\lvert\frac{^1J_{15_{NH}}}{^1J_{14_{NH}}}\right\rvert$ value |
|---|---|---|---|
| ① | 52.4021(15) | $^1J_{14_{NH}}$ | $(2/3)\times\boxed{1}/\textcircled{1}$ |
| ② | 104.7741(10) | $2\times{}^1J_{14_{NH}}$ | $\frac{(2/3)\times\boxed{1}}{(1/2)\times\textcircled{2}}$ |
| ③ | 157.171(8) | $3\times{}^1J_{14_{NH}}$ | $\frac{(2/3)\times\boxed{1}}{(1/3)\times\textcircled{3}}$ |
| $\boxed{1}$ | 110.114(9) | $(3/2)\times{}^1J_{15_{NH}}$ | $\frac{(2/5)\times\boxed{2}}{\textcircled{1}}$ |
| $\boxed{2}$ | 183.554(7) | $(5/2)\times{}^1J_{15_{NH}}$ | $\frac{(2/5)\times\boxed{2}}{(1/2)\times\textcircled{2}}$ |
| | | | $\frac{(2/5)\times\boxed{2}}{(1/3)\times\textcircled{3}}$ |

**Table S2.** Standard Error of the ratio of *J*-coupling frequencies using two different methods. Note that the order of appearance of the $\left\lvert\frac{^1J_{15_{NH}}}{^1J_{14_{NH}}}\right\rvert$ values correspond to the order introduced in the **Table S1**.

| $N_0/n$ | $n$ | $\left\lvert\frac{^1J_{15_{NH}}}{^1J_{14_{NH}}}\right\rvert$ (single-scan fit) | $\left\lvert\frac{^1J_{15_{NH}}}{^1J_{14_{NH}}}\right\rvert$ (statistical mean & error) | $\left\lvert\frac{^1J_{15_{NH}}}{^1J_{14_{NH}}}\right\rvert$ (mean & error from CDF fit) |
|---|---|---|---|---|
| 1☆ | 36000 | 1.4009(7) | N/A | N/A |
| | | 1.40108(18) | | |
| | | 1.40103(14) | | |
| | | 1.4012(6) | | |
| | | 1.40134(15) | | |
| | | 1.40129(9) | | |
| 12 | 3000 | N/A | 1.40100(21) | 1.347(4) |
| | | | 1.40108(5) | 1.40116(4) |
| | | | 1.40102(5) | 1.401034(15) |
| | | | 1.40127(21) | 1.347(4) |
| | | | 1.40135(3) | 1.40139(4) |
| | | | 1.40129(3) | 1.401266(15) |
| 36 | 1000 | N/A | 1.40102(22) | 1.4011(3) |
| | | | 1.40108(4) | 1.401120(24) |
| | | | 1.40102(4) | 1.40109(3) |
| | | | 1.40129(22) | 1.40135(3) |
| | | | 1.40135(3) | 1.401365(22) |
| | | | 1.40129(3) | 1.40133(3) |
| 60 | 600 | N/A | 1.40111(20) | 1.4014(20) |
| | | | 1.40108(3) | 1.401124(24) |
| | | | 1.40102(3) | 1.40093(5) |

| | | | 1.40138(20) | 1.40166(20) |
|---|---|---|---|---|
| | | | 1.401355(25) | 1.401386(20) |
| | | | 1.40129(3) | 1.40119(5) |
| 360 | 100 | N/A | 1.40159(22) | 1.40117(21) |
| | | | 1.401082(23) | 1.401109(21) |
| | | | 1.40102(3) | 1.40105(3) |
| | | | 1.40187(22) | 1.40142(21) |
| | | | 1.401360(20) | 1.401352(20) |
| | | | 1.40129(3) | 1.4013(3) |
| 1000 | 36 | N/A | 1.4028(4) | 1.40133(19) |
| | | | 1.40108(20) | 1.401073(20) |
| | | | 1.40101(3) | 1.40104(3) |
| | | | 1.4031(4) | 1.40159(19) |
| | | | 1.401362(19) | 1.401332(18) |
| | | | 1.40129(3) | 1.40130(3) |
| 2250 | 16 | N/A | 1.4035(5) | 1.40224(20) |
| | | | 1.401075(20) | 1.401071(19) |
| | | | 1.40109(6) | 1.40099(3) |
| | | | 1.4038(5) | 1.40249(20) |
| | | | 1.401361(19) | 1.401326(18) |
| | | | 1.40137(6) | 1.40124(3) |
| 3600 | 10 | N/A | 1.40368(7) | 1.40295(19) |
| | | | 1.401075(20) | 1.401089(19) |
| | | | 1.40193(24) | 1.4001(3) |
| | | | 1.40397(7) | 1.40321(19) |
| | | | 1.401368(19) | 1.401343(19) |
| | | | 1.40223(24) | 1.40125(3) |

**Systematic error analysis procedure**

To test the robustness of the analysis procedure against baseline correction parameters, we studied the variation within the fitting results using different approaches to process the ZULF NMR spectra. **Table S3** is a collection of different reasonable processing steps typically used.

**Table S3**. Different options of the analysis procedure.

| Option | Dropped points (ms) | Removal of moving average | Baseline correction in time domain | Bandpass filter | Zero filling | Phase correction in frequency domain | Baseline correction in frequency domain |
|---|---|---|---|---|---|---|---|
| **#1** | 0.05 | ✓ | ✓ | ✓ | ✓ | ✓ | ✗ |
| **#2** | 0.05 | ✓ | ✓ | ✓ | ✓ | ✓ | ✓ |
| **#3** | 0.05 | ✓ | ✓ | ✓ | ✓ | ✗ | ✓ |
| **#4** | 0.05 | ✓ | ✓ | ✓ | ✗ | ✓ | ✓ |

| #5 | 0.15 | ✓ | ✓ | ✓ | ✓ | ✗ | ✗ |
|---|---|---|---|---|---|---|---|
| #6 | 0.05 | ✓ | ✓ | ✗ | ✓ | ✓ | ✗ |
| #7 | 0.05 | ✓ | ✓ | ✗ | ✓ | ✗ | ✗ |
| #8 | 0.15 | ✓ | ✓ | ✗ | ✓ | ✓ | ✗ |
| #9 | 0.15 | ✓ | ✓ | ✗ | ✓ | ✗ | ✗ |
| #10 | 0.15 | ✓ | ✓ | ✗ | ✗ | ✓ | ✗ |
| #11 | 0.05 | ✓ | ✓ | ✗ | ✗ | ✗ | ✗ |

It is worth noting that options #1-#5 applied the cleaning procedures on the average of all the partitions, whereas in the options #6-11, the baseline of each individual scan was removed before averaging. The different 11 options under consideration yielded different fitted results, shown in **Table S4**.

**Table S4**. Fitted results using different options for the analysis procedure.

| **Statistical Analysis** | | | | | | | | | | | | |
|---|---|---|---|---|---|---|---|---|---|---|---|---|
| Measured $\lvert J_{^{15}NH}/J_{^{14}NH}\rvert$ value | **#1** | **#2** | **#3** | **#4** | **#5** | **#6** | **#7** | **#8** | **#9** | **#10** | **#11** | **Systematic Error (Statistical)** |
| $(4/3)\cdot(\boxed{1}/②)$ | 1.40107 ± 0.000023 | 1.40108 ± 0.000023 | 1.40108 ± 0.000021 | 1.40039 ± 0.000023 | 1.40124 ± 0.000028 | 1.40098 ± 0.000018 | 1.40113 ± 0.000018 | 1.40118 ± 0.000028 | 1.40124 ± 0.000028 | 1.40132 ± 0.000026 | 1.40025 ± 0.000026 | 0.0004 |
| $2\cdot(\boxed{1}/③)$ | 1.40106 ± 0.000032 | 1.40102 ± 0.000031 | 1.40086 ± 0.000025 | 1.40031 ± 0.000053 | 1.40135 ± 0.000033 | 1.40119 ± 0.000031 | 1.40116 ± 0.000025 | 1.40106 ± 0.000042 | 1.40115 ± 0.000035 | 1.40132 ± 0.00026 | 1.4007 ± 0.000026 | 0.0003 |
| $(4/5)\cdot(\boxed{2}/②)$ | 1.4014 ± 0.00002 | 1.40136 ± 0.00002 | 1.4018 ± 0.00002 | 1.40131 ± 0.000026 | 1.40125 ± 0.000029 | 1.40106 ± 0.000018 | 1.40137 ± 0.000019 | 1.40162 ± 0.000027 | 1.40134 ± 0.000028 | 1.40165 ± 0.00004 | 1.40082 ± 0.000032 | 0.0003 |
| $(6/5)\cdot(\boxed{2}/③)$ | 1.40139 ± 0.00003 | 1.40129 ± 0.000029 | 1.40157 ± 0.000024 | 1.40124 ± 0.000055 | 1.40136 ± 0.000034 | 1.40128 ± 0.000031 | 1.4014 ± 0.000025 | 1.4015 ± 0.000042 | 1.40125 ± 0.000035 | 1.40165 ± 0.00027 | 1.40127 ± 0.000032 | 0.00014 |
| **Cumulative Distribution Function fit** | | | | | | | | | | | | |
| Measured $\lvert J_{^{15}NH}/J_{^{14}NH}\rvert$ value | **#1** | **#2** | **#3** | **#4** | **#5** | **#6** | **#7** | **#8** | **#9** | **#10** | **#11** | **Systematic Error (CDF)** |
| $(4/3)\cdot(\boxed{1}/②)$ | 1.40108 ± 0.000022 | 1.4011 ± 0.000022 | 1.4011 ± 0.00002 | 1.40039 ± 0.000023 | 1.40125 ± 0.000027 | 1.40096 ± 0.000017 | 1.40111 ± 0.000017 | 1.40118 ± 0.000028 | 1.40124 ± 0.000028 | 1.40133 ± 0.000025 | 1.40021 ± 0.00002 | 0.0004 |
| $2\cdot(\boxed{1}/③)$ | 1.40107 ± 0.000031 | 1.40103 ± 0.00003 | 1.40087 ± 0.000025 | 1.40023 ± 0.000052 | 1.40135 ± 0.000033 | 1.4012 ± 0.000033 | 1.40116 ± 0.000026 | 1.40107 ± 0.000042 | 1.40117 ± 0.000033 | 1.40108 ± 0.000049 | 1.40067 ± 0.000025 | 0.0003 |
| $(4/5)\cdot(\boxed{2}/②)$ | 1.4014 ± 0.00002 | 1.40136 ± 0.00002 | 1.4018 ± 0.00002 | 1.40133 ± 0.000026 | 1.40125 ± 0.000027 | 1.40104 ± 0.000017 | 1.40135 ± 0.000018 | 1.40162 ± 0.000027 | 1.40133 ± 0.000028 | 1.40164 ± 0.00004 | 1.40077 ± 0.000028 | 0.0003 |
| $(6/5)\cdot(\boxed{2}/③)$ | 1.40139 ± 0.00003 | 1.4013 ± 0.000029 | 1.40157 ± 0.000024 | 1.40117 ± 0.000054 | 1.40135 ± 0.000033 | 1.40127 ± 0.000033 | 1.4014 ± 0.000026 | 1.4015 ± 0.000041 | 1.40126 ± 0.000033 | 1.40138 ± 0.000059 | 1.40122 ± 0.000032 | 0.00012 |

Since the results presented in the **Table S4** differ between the options well beyond the error bars of the individual values of the ratio, we construct in the last column a "systematic error" defined as the standard deviation of the results taking these 11 different options of data processing. These systematic errors are the error bars shown in **Figure 2c** of the main text.

As an example, **Figure S3** shows the results for options #2, #6, and #11, respectively, applied on 360 partitions of 100 averaged scans each. The black dashed lines indicate the values obtained by direct fit of all 360 scans, and the red dashed lines indicate the value that was obtained with the data collected in Berkeley using a homebuilt zero-field spectrometer.[7]

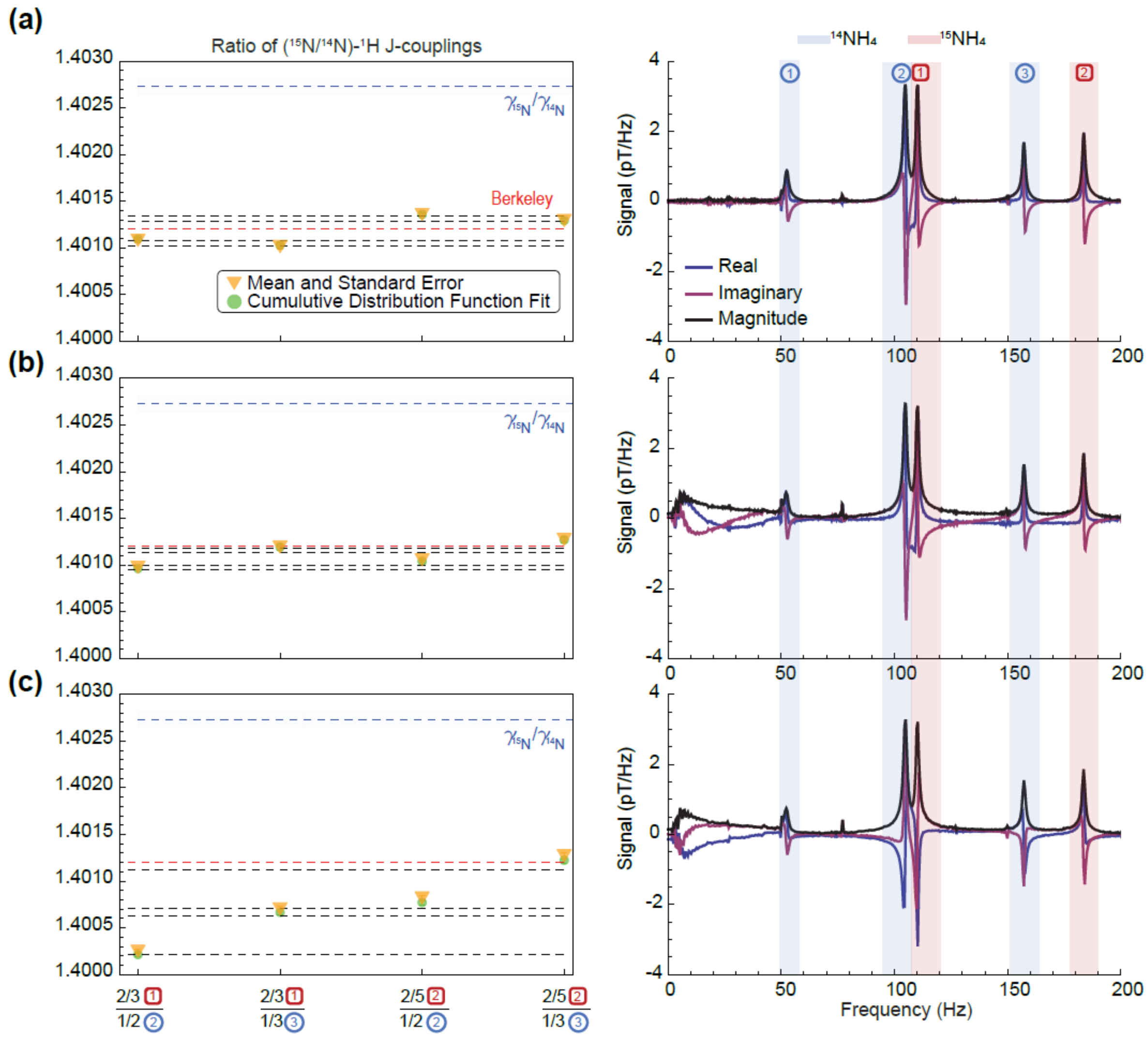

**Figure S3.** Results showing the estimated $\left|J_{^{15}\mathrm{NH}}/J_{^{14}\mathrm{NH}}\right|$ ratio using **(a)** option #2, **(b)** option #8, and **(c)** option #11 of the postprocessing procedure.

## B. Pulse-length dependence

To investigate opportunities of resolving $^{1}$H-D $J$-coupling with zero-field NMR techniques, we simulated the zero-field spectra for deuterated ammonium isotopologues and analyzed their energy level structures using perturbation theory.

For all isotopologues $^{15}\mathrm{ND}_x\mathrm{H}^{+}_{4-x}$ (where $x = 1 - 3$) one can distinguish regions of low- and high-frequency peaks (**Figure S4a**). High-frequency peaks correspond to transitions within the strongly-coupled subsystem consisting of $^{15}$N and $^{1}$H spins when total deuterium spin remains unchanged. Low-frequency peaks correspond to the deuterium spin flips keeping the strongly-coupled $^{15}$N-$^{1}$H subsystem unperturbed.

These groups of peaks are expected to respond differently to the magnetic pulse excitation and thus, by plotting their integrals as a function of magnetic pulse length, one can extract information about subtle spin-spin interactions which otherwise would have taken more sophisticated multinuclear high-field NMR analysis.

For example, for $^{15}NDH_3^+$, one can clearly see that different $^1$H-D *J*-couplings result in different calculated dependencies of low-frequency integrals on the excitation pulse length (**Figure S4b**). Even a subtle change of ~3 Hz in $J_{HD}$ dramatically modifies the pulse length dependence. Importantly, not only a magnitude but also a sign of the *J*-coupling can be understood by experimentally probing the pulse-length dependence (**Figure 5**).

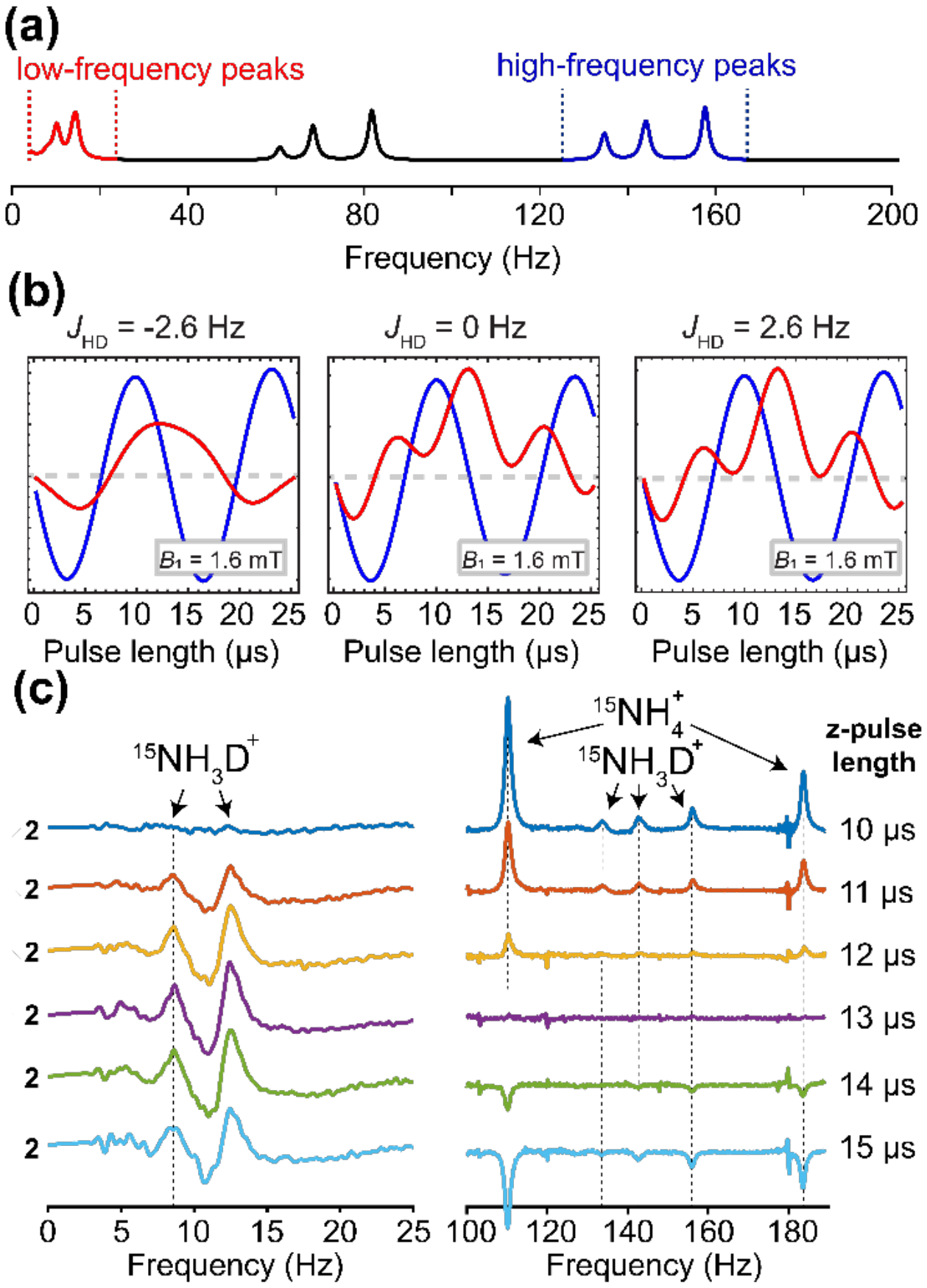

**Figure S4**. (a) Calculated zero-field *J*-spectrum of $^{15}NDH_3^+$ featuring low-frequency (0-15 Hz, red) and high-frequency (120-180 Hz, blue) regions of the spectrum. (b) Calculated integrals for the low-frequency (0-15 Hz, red) and high-frequency (120-180 Hz, blue) peaks in the spectrum of $^{15}NDH_3^+$ as a function of the magnetic pulse excitation length assuming $J_{HD}$ = -2.6 Hz (left), 0 Hz (middle), 2.6 Hz (right). (c) ZULF-NMR spectra of $^{15}ND_xH_{4-x}^+$ solution with deuterium fraction $p$ = 24% recorded after the action of constant-amplitude magnetic-field pulses with various duration (10-15 μs, amplitude 1.6 mT) applied in the direction of the magnetometer-detection axis. Note that the low-frequency (0-15 Hz) and high-frequency (100-200 Hz) parts of the spectrum respond differently to the magnetic-pulse excitation.

## C. $T_1$ measurement of $^{15}$N ammonia

The $T_1$ measurement was performed in a Magritek Spinsolve 1.1 T on a solution of 6 M ammonium chloride with 50% $^{15}$N isotopic labeling dissolved in aqueous sulfuric acid (1.9 M). The proton $T_1$ measurement was carried out using a standard inversion-recovery experiment for the two different ammonium isotopologues (**Figure S5a-b**). The parameters for the experiment were: four scans, 6.4 s of acquisition, 2 min delay between scans, 10 s maximum inversion time, four dummy scans, and 27 steps. A refocused INEPT (insensitive nuclei enhanced by polarization transfer) pulse sequence was used to transfer polarization from the protons of $NH_4^+$ to $^{15}$N using an interpulse delay of $1/(4J)$ ($|J|$= 74 Hz) and a refocus delay of 1.135 ms. A 90° pulse was applied to convert the coherence generated in the refocused INEPT experiment back into $S_z$ spin order in $^{15}$N which was then allowed to relax for a variable time at 1 tesla before detection. The measured points represent integrals taken from $^{15}$N spectra that are the sum of 16 averages with an interscan delay of 20 s (**Figure S5c**). The estimated $T_1 = (49 \pm 3)$ s is the result of fitting with a mono-exponential decay function and error is estimated as the standard error of the fit.

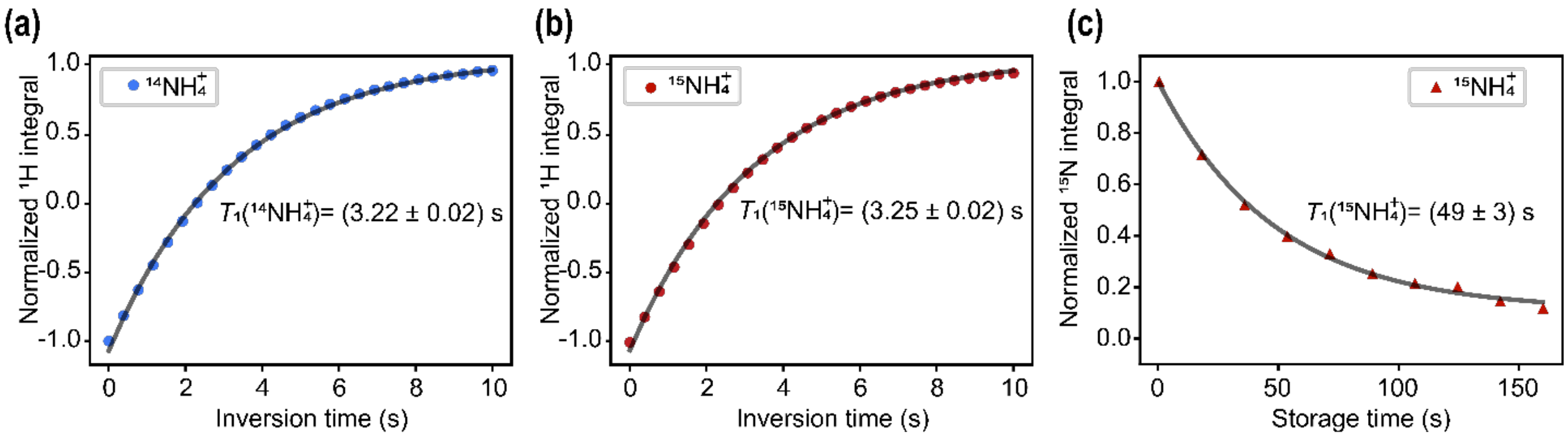

**Figure S5.** (a) Decay of an $^1$H signal as a function of inversion time measured for $^{14}$N-ammonium. (b) Inversion-recovery measurement for $^{15}$N-ammonium. (c) Decay of an $^{15}$N signal as a function of storage time measured using refocused INEPT.

## D. Analytical calculations

Amongst all possible nuclear spin interactions, only electron-mediated *J*-couplings are important for analyzing the zero-field NMR spectra of molecules in the liquid state. The nuclear spin Hamiltonian ($\hat{H}$) for different isotopologues of the ammonium cation ($^{15}ND_xH_{4-x}^+$, where $x$ can take values from 0 to 4) is therefore given by

$$\hat{H} = J_{\mathrm{NH}}(\hat{\mathbf{S}} \cdot \hat{\mathbf{K}}_{\mathrm{A}}) + J_{\mathrm{ND}}(\hat{\mathbf{S}} \cdot \hat{\mathbf{K}}_{\mathrm{B}}) + J_{\mathrm{HD}}(\hat{\mathbf{K}}_{\mathrm{A}} \cdot \hat{\mathbf{K}}_{\mathrm{B}})\,, \tag{S3}$$

where we follow the notation suggested by Butler et al.,[4] i.e., $\hat{\mathbf{S}}$ denotes the spin of the heteronucleus (in our case, $^{15}$N or $^{14}$N), $\hat{\mathbf{K}}_{\mathrm{A}}$ denotes the total proton spin ($\hat{\mathbf{K}}_{\mathrm{A}} = \sum_{i=1}^{4-x} \hat{\mathbf{I}}_i^{\mathrm{H}}$), and $\hat{\mathbf{K}}_{\mathrm{B}}$ denotes the total deuterium spin ($\hat{\mathbf{K}}_{\mathrm{B}} = \sum_{i=1}^{x} \hat{\mathbf{I}}_i^{\mathrm{D}}$). The values of the couplings were extracted from the experimentally measured spectra and are shown in the **Table S5**. Notice that due to the electronic similarity of $^1$H and D, the coupling between $^{15}$N and D was estimated as $J_{\mathrm{ND}} \approx J_{\mathrm{NH}} \left(\frac{\gamma_{\mathrm{D}}}{\gamma_{\mathrm{H}}}\right)$, ignoring isotope effects. Below we derive the zero-field NMR frequencies in *J*-spectra for various isotopologues of ammonium cation.

**Table S5.** Heteronuclear $J$-couplings in the $^{15}ND_xH_{4-x}^+$ and $^{14}NH_4^+$ spin systems measured in this work.

| $J$-coupling | Associated peaks | Extracted coupling value (Hz) |
|---|---|---|
| $^1J_{^{15}NH}$ * | $(2/3) \times \boxed{1}$ | -73.410(6) |
| | $(2/5) \times \boxed{2}$ | -73.422(3) |
| $^1J_{^{14}NH}$ * | ① | 52.4021(15) |
| | $(1/2) \times$ ② | 52.392(5) |
| | $(1/3) \times$ ③ | 52.390(3) |
| $^1J_{^{15}ND}$ ** | | -11.3(1) |
| $^2J_{HD}$ * | | -2.6(1) |

*data extracted from the fit of averaged 36000 zero-field NMR spectra;

**data extracted from the high-field (800 MHz) $^{15}$N NMR spectra.

### a. Zero-field NMR spectrum of $^{15}NH_4^+$

For $x = 0$ Eq. (S3) simplifies to

$$\hat{H} = J_{NH}(\hat{\mathbf{S}} \cdot \hat{\mathbf{K}}_A) . \tag{S4}$$

By introducing the total spin $\hat{\mathbf{F}}_A$ of the $^{15}$N-$^1$H system as $\hat{\mathbf{F}}_A = \hat{\mathbf{S}} + \hat{\mathbf{K}}_A$, one can show that

$$\hat{\mathbf{S}} \cdot \hat{\mathbf{K}}_A = \frac{1}{2}\left(\hat{\mathbf{F}}_A^2 - \hat{\mathbf{S}}^2 - \hat{\mathbf{K}}_A^2\right) . \tag{S5}$$

Thus, since the value of $S$ equals to ½, eigenstates of the Hamiltonian (Eq. S4) denoted $|F_A, K_A\rangle$, can be expressed using quantum numbers $F_A$ and $K_A$ (corresponding to the operators $\hat{\mathbf{F}}_A^2$ and $\hat{\mathbf{K}}_A^2$) and their energies ($E$) can be determined using the equation

$$E = \frac{J_{NH}}{2}\left(F_A(F_A + 1) - S(S + 1) - K_A(K_A + 1)\right) . \tag{S6}$$

Since $\hat{\mathbf{K}}_A^2$ and $\hat{\mathbf{S}}^2$ commute with the perturbation (magnetic DC pulse along $z$-direction), transitions between eigenstates are constrained by $\Delta K_A = \Delta S = 0$, while $\Delta F_A = 0, \pm 1$. Therefore, for $^{15}NH_4^+$ we expect peaks at $\frac{5}{2}J_{NH}$ and $\frac{3}{2}J_{NH}$, corresponding to transitions between $\left|\frac{5}{2}, 2\right\rangle \rightarrow \left|\frac{3}{2}, 2\right\rangle$ and $\left|\frac{3}{2}, 1\right\rangle \rightarrow \left|\frac{1}{2}, 1\right\rangle$ as shown in **Figure 1b** and **Table S6**.

**Table S6.** Nuclear spin eigenstates, corresponding energies, allowed transitions and zero-field NMR spectral frequencies of $^{15}NH_4^+$ ion.

| Eigenstates, $|F_A, K_A\rangle$ | Energy | Transition | Frequency |
|---|---|---|---|
| $\left|\frac{5}{2}, 2\right\rangle$ | $J_{NH}$ | $\left|\frac{5}{2}, 2\right\rangle \rightarrow \left|\frac{3}{2}, 2\right\rangle$ | $\frac{5}{2}J_{NH}$ |

| $\left\|\frac{3}{2},2\right\rangle$ | $-\frac{3}{2}J_{\mathrm{NH}}$ | $\left\|\frac{3}{2},1\right\rangle \rightarrow \left\|\frac{1}{2},1\right\rangle$ | $\frac{3}{2}J_{\mathrm{NH}}$ |
|---|---|---|---|
| $\left\|\frac{3}{2},1\right\rangle$ | $\frac{1}{2}J_{\mathrm{NH}}$ | | |
| $\left\|\frac{1}{2},1\right\rangle$ | $-J_{\mathrm{NH}}$ | | |
| $\left\|\frac{1}{2},0\right\rangle$ | 0 | | |

### b. Zero-field NMR spectrum of $^{15}\mathrm{NDH}_3^+$

Since $|J_{\mathrm{ND}}|, |J_{\mathrm{HD}}| < |J_{\mathrm{NH}}|$, we can apply perturbation theory and split the Hamiltonian (S4) into two parts:

$$\hat{H} = \hat{H}_0 + \hat{H}_1\,, \tag{S7}$$

where $\hat{H}_0 = J_{\mathrm{NH}}(\hat{\mathbf{S}} \cdot \hat{\mathbf{K}}_{\mathrm{A}})$ is the same as in Eq. (S4) and $\hat{H}_1 = J_{\mathrm{ND}}(\hat{\mathbf{S}} \cdot \hat{\mathbf{K}}_{\mathrm{B}}) + J_{\mathrm{HD}}(\hat{\mathbf{K}}_{\mathrm{A}} \cdot \hat{\mathbf{K}}_{\mathrm{B}})$. Hamiltonian $\hat{H}_1$ can be simplified further. Using geometric arguments[4], one may show that $\hat{\mathbf{S}}$ and $\hat{\mathbf{K}}_{\mathrm{A}}$ can be replaced by their projections on $\hat{\mathbf{F}}_{\mathrm{A}}$:

$$\hat{H}_1 = J_{\mathrm{ND}}(\hat{\mathbf{S}} \cdot \hat{\mathbf{K}}_{\mathrm{B}}) + J_{\mathrm{HD}}(\hat{\mathbf{K}}_{\mathrm{A}} \cdot \hat{\mathbf{K}}_{\mathrm{B}}) = J_{\mathrm{ND}}\frac{\langle \hat{\mathbf{S}} \cdot \hat{\mathbf{F}}_{\mathrm{A}} \rangle}{\langle \hat{\mathbf{F}}_{\mathrm{A}}^2 \rangle}(\hat{\mathbf{F}}_{\mathrm{A}} \cdot \hat{\mathbf{K}}_{\mathrm{B}}) + J_{\mathrm{HD}}\frac{\langle \hat{\mathbf{K}}_{\mathrm{A}} \cdot \hat{\mathbf{F}}_{\mathrm{A}} \rangle}{\langle \hat{\mathbf{F}}_{\mathrm{A}}^2 \rangle}(\hat{\mathbf{F}}_{\mathrm{A}} \cdot \hat{\mathbf{K}}_{\mathrm{B}}) =$$

$$= \frac{J_{\mathrm{ND}}^{\parallel} + J_{\mathrm{HD}}^{\parallel}}{2}\left(\hat{\mathbf{F}}^2 - \hat{\mathbf{F}}_{\mathrm{A}}^2 - \hat{\mathbf{K}}_{\mathrm{B}}^2\right), \tag{S8}$$

where $\hat{\mathbf{F}} = \hat{\mathbf{F}}_{\mathrm{A}} + \hat{\mathbf{K}}_{\mathrm{B}}$ is the total spin of the system. This notation suggests that there are no longer single couplings between protons and the heteronucleus to the deuterium atoms but rather a total coupling of the strongly coupled system $(\hat{\mathbf{S}}, \hat{\mathbf{K}}_{\mathrm{A}})$ to the weakly coupled deuterium $(\hat{\mathbf{K}}_{\mathrm{B}})$. Since $\hat{\mathbf{K}}_{\mathrm{A}} = \hat{\mathbf{F}}_{\mathrm{A}} - \hat{\mathbf{S}}$ and $\hat{\mathbf{S}} = \hat{\mathbf{F}}_{\mathrm{A}} - \hat{\mathbf{K}}_{\mathrm{A}}$, by analogy with (S5) one may find that

$$J_{\mathrm{ND}}^{\parallel} = J_{\mathrm{ND}}\frac{\langle \hat{\mathbf{S}} \cdot \hat{\mathbf{F}}_{\mathrm{A}} \rangle}{\langle \hat{\mathbf{F}}_{\mathrm{A}}^2 \rangle} = J_{\mathrm{ND}}\frac{\left(F_{\mathrm{A}}(F_{\mathrm{A}}+1) + S(S+1) - K_{\mathrm{A}}(K_{\mathrm{A}}+1)\right)}{2F_{\mathrm{A}}(F_{\mathrm{A}}+1)}\,, \tag{S9}$$

and

$$J_{\mathrm{HD}}^{\parallel} = J_{\mathrm{HD}}\frac{\langle \hat{\mathbf{K}}_{\mathrm{A}} \cdot \hat{\mathbf{F}}_{\mathrm{A}} \rangle}{\langle \hat{\mathbf{F}}_{\mathrm{A}}^2 \rangle} = J_{\mathrm{HD}}\frac{\left(F_{\mathrm{A}}(F_{\mathrm{A}}+1) + K_{\mathrm{A}}(K_{\mathrm{A}}+1) - S(S+1)\right)}{2F_{\mathrm{A}}(F_{\mathrm{A}}+1)}\,. \tag{S10}$$

Since $K_{\mathrm{B}} = 1$, i.e., it is fixed, and we are left with $F$, $F_{\mathrm{A}}$, and $K_{\mathrm{A}}$ as the quantum numbers for defining the eigenstates of $\hat{H}$ which will be denoted as $|F, F_{\mathrm{A}}, K_{\mathrm{A}}\rangle$. **Table S7** shows eigenstates, allowed transitions and corresponding spectral lines. Note that we distinguish between high-frequency and low-frequency transitions. High-frequency transitions are transitions in which the quantum number of the strongly coupled system ($F_{\mathrm{A}}$) changes. If $\Delta F_{\mathrm{A}} = 0$ and only $F$ changes, we will refer to these transitions as low-frequency transitions. These transitions can therefore be seen as deuterium spin flip. The only allowed transitions are those that leave $K_{\mathrm{A}}$ unchanged.

**Table S7.** Nuclear spin eigenstates, corresponding energies, allowed transitions and zero-field NMR spectral frequencies of $^{15}NDH_3^+$ ion ($K_B$=1 and it is fixed).

| Eigenstate, $\|F, F_A, K_A\rangle$ | $E_0$ | $E_1$ |
|---|---|---|
| $\left\|3, 2, \frac{3}{2}\right\rangle$ | $\frac{3}{4}J_{NH}$ | $\frac{1}{2}(J_{ND} + 3J_{HD})$ |
| $\left\|2, 2, \frac{3}{2}\right\rangle$ | $\frac{3}{4}J_{NH}$ | $-\frac{1}{4}(J_{ND} + 3J_{HD})$ |
| $\left\|1, 2, \frac{3}{2}\right\rangle$ | $\frac{3}{4}J_{NH}$ | $-\frac{3}{4}(J_{ND} + 3J_{HD})$ |
| $\left\|2, 1, \frac{3}{2}\right\rangle$ | $-\frac{5}{4}J_{NH}$ | $\frac{1}{4}(-J_{ND} + 5J_{HD})$ |
| $\left\|1, 1, \frac{3}{2}\right\rangle$ | $-\frac{5}{4}J_{NH}$ | $-\frac{1}{4}(-J_{ND} + 5J_{HD})$ |
| $\left\|0, 1, \frac{3}{2}\right\rangle$ | $-\frac{5}{4}J_{NH}$ | $-\frac{1}{2}(-J_{ND} + 5J_{HD})$ |
| $\left\|2, 1, \frac{1}{2}\right\rangle$ | $\frac{1}{4}J_{NH}$ | $\frac{1}{2}(J_{ND} + J_{HD})$ |
| $\left\|1, 1, \frac{1}{2}\right\rangle$ | $\frac{1}{4}J_{NH}$ | $-\frac{1}{2}(J_{ND} + J_{HD})$ |
| $\left\|0, 1, \frac{1}{2}\right\rangle$ | $\frac{1}{4}J_{NH}$ | $-(J_{ND} + J_{HD})$ |
| $\left\|1, 0, \frac{1}{2}\right\rangle$ | $-\frac{3}{4}J_{NH}$ | $0$ |

| Transition | Frequency (high) |
|---|---|
| $\left\|3, 2, \frac{3}{2}\right\rangle \rightarrow \left\|2, 1, \frac{3}{2}\right\rangle$ | $2J_{NH} + \frac{3}{4}J_{ND} + \frac{1}{4}J_{HD}$ |
| $\left\|2, 2, \frac{3}{2}\right\rangle \rightarrow \left\|2, 1, \frac{3}{2}\right\rangle$ | $2J_{NH} - 2J_{HD}$ |
| $\left\|2, 2, \frac{3}{2}\right\rangle \rightarrow \left\|1, 1, \frac{3}{2}\right\rangle$ | $2J_{NH} - \frac{1}{2}J_{ND} + \frac{1}{2}J_{HD}$ |
| $\left\|1, 2, \frac{3}{2}\right\rangle \rightarrow \left\|2, 1, \frac{3}{2}\right\rangle$ | $2J_{NH} - \frac{1}{2}J_{ND} - \frac{7}{2}J_{HD}$ |
| $\left\|1, 2, \frac{3}{2}\right\rangle \rightarrow \left\|1, 1, \frac{3}{2}\right\rangle$ | $2J_{NH} - J_{ND} - J_{HD}$ |
| $\left\|1, 2, \frac{3}{2}\right\rangle \rightarrow \left\|0, 1, \frac{3}{2}\right\rangle$ | $2J_{\mathrm{NH}} - \frac{5}{4}J_{\mathrm{ND}} + \frac{1}{4}J_{\mathrm{HD}}$ |
| $\left\|2, 1, \frac{1}{2}\right\rangle \rightarrow \left\|1, 0, \frac{1}{2}\right\rangle$ | $J_{\mathrm{NH}} + \frac{1}{2}J_{\mathrm{ND}} + \frac{1}{2}J_{\mathrm{HD}}$ |
| $\left\|1, 1, \frac{1}{2}\right\rangle \rightarrow \left\|1, 0, \frac{1}{2}\right\rangle$ | $J_{\mathrm{NH}} - \frac{1}{2}J_{\mathrm{ND}} - \frac{1}{2}J_{\mathrm{HD}}$ |
| $\left\|0, 1, \frac{1}{2}\right\rangle \rightarrow \left\|1, 0, \frac{1}{2}\right\rangle$ | $2J_{\mathrm{NH}} - J_{\mathrm{ND}} - J_{\mathrm{HD}}$ |

| Transition | Frequency (low) |
|---|---|
| $\left\|3, 2, \frac{3}{2}\right\rangle \rightarrow \left\|2, 2, \frac{3}{2}\right\rangle$ | $\frac{3}{4}J_{\mathrm{ND}} + \frac{9}{4}J_{\mathrm{HD}}$ |
| $\left\|2, 2, \frac{3}{2}\right\rangle \rightarrow \left\|1, 2, \frac{3}{2}\right\rangle$ | $\frac{1}{2}J_{\mathrm{ND}} + \frac{3}{2}J_{\mathrm{HD}}$ |
| $\left\|1, 1, \frac{3}{2}\right\rangle \rightarrow \left\|0, 1, \frac{3}{2}\right\rangle$ | $-\frac{1}{4}J_{\mathrm{ND}} + \frac{5}{4}J_{\mathrm{HD}}$ |
| $\left\|1, 1, \frac{1}{2}\right\rangle \rightarrow \left\|0, 1, \frac{1}{2}\right\rangle$ | $\frac{1}{2}J_{\mathrm{ND}} + \frac{1}{2}J_{\mathrm{HD}}$ |

### c. Zero-field NMR spectrum of $^{15}\mathrm{ND_2H_2^+}$

We use the same approach for calculating nuclear energy levels of the ion $^{15}\mathrm{ND_2H_2^+}$. However, here we have an additional degree of freedom since $K_{\mathrm{B}}$ can now take values 0, 1 and 2. States are denoted $|F, F_{\mathrm{A}}, K_{\mathrm{A}}\rangle$ and $K_{\mathrm{B}}$ is given explicitly (**Table S8**). One should note that perturbation theory will eventually break down for states with larger $K_{\mathrm{B}}$ (specifically, $K_{\mathrm{B}} = 2$ in this case) since the corresponding term in the Hamiltonian increases. However, states with $K_{\mathrm{B}} = 1$ are still approximated well by perturbation theory. The selection rules introduced above remain unchanged.

**Table S8.** Nuclear spin eigenstates, corresponding energies, allowed transitions and zero-field NMR spectral frequencies of $^{15}\mathrm{ND_2H_2^+}$ ion.

| $K_{\mathrm{B}}$ | Eigenstate, $\|F, F_{\mathrm{A}}, K_{\mathrm{A}}\rangle$ | $E_0$ | $E_1$ |
|---|---|---|---|
| 0 | $\left\|\frac{3}{2}, \frac{3}{2}, 1\right\rangle$ | $\frac{1}{2}J_{\mathrm{NH}}$ | 0 |
| | $\left\|\frac{1}{2}, \frac{1}{2}, 0\right\rangle$ | 0 | 0 |
| 1 | $\left\|\frac{5}{2}, \frac{3}{2}, 1\right\rangle$ | $\frac{1}{2}J_{\mathrm{NH}}$ | $\frac{1}{2}\left(J_{\mathrm{ND}} + 2J_{\mathrm{HD}}\right)$ |
| | $\left\|\frac{3}{2},\frac{3}{2},1\right\rangle$ | $\frac{1}{2}J_{\mathrm{NH}}$ | $-\frac{1}{3}(J_{\mathrm{ND}}+2J_{\mathrm{HD}})$ |
| | $\left\|\frac{1}{2},\frac{3}{2},1\right\rangle$ | $\frac{1}{2}J_{\mathrm{NH}}$ | $-\frac{5}{6}(J_{\mathrm{ND}}+2J_{\mathrm{HD}})$ |
| | $\left\|\frac{3}{2},\frac{1}{2},1\right\rangle$ | $-J_{\mathrm{NH}}$ | $-\frac{1}{6}(J_{\mathrm{ND}}-4J_{\mathrm{HD}})$ |
| | $\left\|\frac{1}{2},\frac{1}{2},1\right\rangle$ | $-J_{\mathrm{NH}}$ | $\frac{1}{3}(J_{\mathrm{ND}}-4J_{\mathrm{HD}})$ |
| | $\left\|\frac{3}{2},\frac{1}{2},0\right\rangle$ | 0 | $\frac{1}{2}J_{\mathrm{ND}}$ |
| | $\left\|\frac{1}{2},\frac{1}{2},0\right\rangle$ | 0 | $-J_{\mathrm{ND}}$ |
| 2 | $\left\|\frac{7}{2},\frac{3}{2},1\right\rangle$ | $\frac{1}{2}J_{\mathrm{NH}}$ | $J_{\mathrm{ND}}+2J_{\mathrm{HD}}$ |
| | $\left\|\frac{5}{2},\frac{3}{2},1\right\rangle$ | $\frac{1}{2}J_{\mathrm{NH}}$ | $-\frac{1}{6}(J_{\mathrm{ND}}+2J_{\mathrm{HD}})$ |
| | $\left\|\frac{3}{2},\frac{3}{2},1\right\rangle$ | $\frac{1}{2}J_{\mathrm{NH}}$ | $-(J_{\mathrm{ND}}+2J_{\mathrm{HD}})$ |
| | $\left\|\frac{1}{2},\frac{3}{2},1\right\rangle$ | $\frac{1}{2}J_{\mathrm{NH}}$ | $-\frac{3}{2}(J_{\mathrm{ND}}+2J_{\mathrm{HD}})$ |
| | $\left\|\frac{5}{2},\frac{1}{2},1\right\rangle$ | $-J_{\mathrm{NH}}$ | $-\frac{1}{3}(J_{\mathrm{ND}}-4J_{\mathrm{HD}})$ |
| | $\left\|\frac{3}{2},\frac{1}{2},1\right\rangle$ | $-J_{\mathrm{NH}}$ | $\frac{1}{2}(J_{\mathrm{ND}}-4J_{\mathrm{HD}})$ |
| | $\left\|\frac{1}{2},\frac{1}{2},1\right\rangle$ | $-J_{\mathrm{NH}}$ | $(J_{\mathrm{ND}}+2J_{\mathrm{HD}})$ |
| | $\left\|\frac{5}{2},\frac{1}{2},0\right\rangle$ | 0 | $J_{\mathrm{ND}}$ |
| | $\left\|\frac{3}{2},\frac{1}{2},0\right\rangle$ | 0 | $-\frac{3}{2}J_{\mathrm{ND}}$ |

| $K_{\mathrm{B}}$ | Transition | Frequency (high) |
|---|---|---|
| 1 | $\left\|\frac{5}{2},\frac{3}{2},1\right\rangle \rightarrow \left\|\frac{3}{2},\frac{1}{2},1\right\rangle$ | $\frac{3}{2}J_{\mathrm{NH}}+\frac{2}{3}J_{\mathrm{ND}}+\frac{1}{3}J_{\mathrm{HD}}$ |
| | $\left\|\frac{3}{2},\frac{3}{2},1\right\rangle \rightarrow \left\|\frac{3}{2},\frac{1}{2},1\right\rangle$ | $\frac{3}{2}J_{\mathrm{NH}}-\frac{1}{6}J_{\mathrm{ND}}-\frac{4}{3}J_{\mathrm{HD}}$ |
| | $\left\|\frac{3}{2},\frac{3}{2},1\right\rangle \rightarrow \left\|\frac{1}{2},\frac{1}{2},1\right\rangle$ | $\frac{3}{2}J_{\mathrm{NH}}-\frac{2}{3}J_{\mathrm{ND}}+\frac{2}{3}J_{\mathrm{HD}}$ |
| | $\left\|\frac{1}{2},\frac{3}{2},1\right\rangle \rightarrow \left\|\frac{1}{2},\frac{1}{2},1\right\rangle$ | $\frac{3}{2}J_{\mathrm{NH}}-\frac{7}{6}J_{\mathrm{ND}}-\frac{1}{3}J_{\mathrm{HD}}$ |
| 2 | $\left\|\frac{7}{2},\frac{3}{2},1\right\rangle \rightarrow \left\|\frac{5}{2},\frac{1}{2},1\right\rangle$ | $\frac{3}{2}J_{\mathrm{NH}}+\frac{4}{3}J_{\mathrm{ND}}+\frac{2}{3}J_{\mathrm{HD}}$ |
| | $\left\|\frac{5}{2},\frac{3}{2},1\right\rangle \rightarrow \left\|\frac{5}{2},\frac{1}{2},1\right\rangle$ | $\frac{3}{2}J_{\mathrm{NH}}+\frac{1}{6}J_{\mathrm{ND}}-\frac{5}{3}J_{\mathrm{HD}}$ |
| | $\left\|\frac{5}{2},\frac{3}{2},1\right\rangle \rightarrow \left\|\frac{3}{2},\frac{1}{2},1\right\rangle$ | $\frac{3}{2}J_{\mathrm{NH}}-\frac{2}{3}J_{\mathrm{ND}}+\frac{5}{3}J_{\mathrm{HD}}$ |
| | $\left\|\frac{3}{2},\frac{3}{2},1\right\rangle \rightarrow \left\|\frac{5}{2},\frac{1}{2},1\right\rangle$ | $\frac{3}{2}J_{\mathrm{NH}}-\frac{2}{3}J_{\mathrm{ND}}-\frac{10}{3}J_{\mathrm{HD}}$ |
| | $\left\|\frac{3}{2},\frac{3}{2},1\right\rangle \rightarrow \left\|\frac{3}{2},\frac{1}{2},1\right\rangle$ | $\frac{3}{2}J_{\mathrm{NH}}-\frac{3}{2}J_{\mathrm{ND}}$ |
| | $\left\|\frac{3}{2},\frac{3}{2},1\right\rangle \rightarrow \left\|\frac{1}{2},\frac{1}{2},1\right\rangle$ | $\frac{3}{2}J_{\mathrm{NH}}-2J_{\mathrm{ND}}+2J_{\mathrm{HD}}$ |
| | $\left\|\frac{1}{2},\frac{3}{2},1\right\rangle \rightarrow \left\|\frac{3}{2},\frac{1}{2},1\right\rangle$ | $\frac{3}{2}J_{\mathrm{NH}}-2J_{\mathrm{ND}}-J_{\mathrm{HD}}$ |
| | $\left\|\frac{1}{2},\frac{3}{2},1\right\rangle \rightarrow \left\|\frac{1}{2},\frac{1}{2},1\right\rangle$ | $\frac{3}{2}J_{\mathrm{NH}}-\frac{5}{2}J_{\mathrm{ND}}+J_{\mathrm{HD}}$ |

| $K_{\mathrm{B}}$ | Transition | Spectral frequency (low) |
|---|---|---|
| 1 | $\left\|\frac{5}{2},\frac{3}{2},1\right\rangle \rightarrow \left\|\frac{3}{2},\frac{3}{2},1\right\rangle$ | $\frac{5}{6}J_{\mathrm{ND}}+\frac{5}{3}J_{\mathrm{HD}}$ |
| | $\left\|\frac{3}{2},\frac{3}{2},1\right\rangle \rightarrow \left\|\frac{1}{2},\frac{3}{2},1\right\rangle$ | $\frac{1}{2}J_{\mathrm{ND}}+J_{\mathrm{HD}}$ |
| | $\left\|\frac{3}{2},\frac{1}{2},1\right\rangle \rightarrow \left\|\frac{1}{2},\frac{1}{2},1\right\rangle$ | $-\frac{1}{2}J_{\mathrm{ND}}+2J_{\mathrm{HD}}$ |
| | $\left\|\frac{3}{2},\frac{1}{2},0\right\rangle \rightarrow \left\|\frac{1}{2},\frac{1}{2},0\right\rangle$ | $\frac{3}{2}J_{\mathrm{ND}}$ |
| 2 | $\left\|\frac{7}{2},\frac{3}{2},1\right\rangle \rightarrow \left\|\frac{5}{2},\frac{3}{2},1\right\rangle$ | $\frac{7}{6}J_{\mathrm{ND}}+\frac{7}{3}J_{\mathrm{HD}}$ |
| | $\left\|\frac{5}{2},\frac{3}{2},1\right\rangle \rightarrow \left\|\frac{3}{2},\frac{3}{2},1\right\rangle$ | $\frac{5}{6}J_{\mathrm{ND}}+\frac{5}{3}J_{\mathrm{HD}}$ |
| | $\left\|\frac{3}{2},\frac{3}{2},1\right\rangle \rightarrow \left\|\frac{1}{2},\frac{3}{2},1\right\rangle$ | $\frac{1}{2}J_{\mathrm{ND}}+J_{\mathrm{HD}}$ |
| | $\left\|\frac{5}{2},\frac{1}{2},1\right\rangle \rightarrow \left\|\frac{3}{2},\frac{1}{2},1\right\rangle$ | $-\frac{5}{6}J_{\mathrm{ND}}+\frac{10}{3}J_{\mathrm{HD}}$ |
| | $\left\|\frac{3}{2},\frac{1}{2},1\right\rangle \rightarrow \left\|\frac{1}{2},\frac{1}{2},1\right\rangle$ | $-\frac{1}{2}J_{\mathrm{ND}}+2J_{\mathrm{HD}}$ |

| $\left\|\frac{5}{2},\frac{1}{2},0\right\rangle \rightarrow \left\|\frac{3}{2},\frac{1}{2},0\right\rangle$ | $\frac{5}{2}J_{\mathrm{ND}}$ |
|---|---|

## d. Zero-field NMR spectrum of $^{15}\mathrm{NHD}_3^+$

Analogously, we derive energy levels and corresponding transitions for the $^{15}\mathrm{ND}_3\mathrm{H}^+$ ion (**Table S9**). One may note that actual frequencies may deviate from the derived ones due to the fact that perturbation theory no longer holds for large $K_\mathrm{B}$ values.

**Table S9.** Nuclear spin eigenstates, corresponding energies, allowed transitions and zero-field NMR spectral frequencies of $^{15}\mathrm{ND}_3\mathrm{H}^+$ ion.

| $K_\mathrm{B}$ | Eigenstate, $\|F,K_\mathrm{B},F_\mathrm{A}\rangle$ | $E_0$ | $E_1$ |
|---|---|---|---|
| 0 | $\|4,3,1\rangle$ | $\frac{1}{4}J_{\mathrm{NH}}$ | $\frac{3}{2}J_{\mathrm{ND}}+\frac{3}{2}J_{\mathrm{HD}}$ |
| | $\|3,3,1\rangle$ | $\frac{1}{4}J_{\mathrm{NH}}$ | $-\frac{1}{2}J_{\mathrm{ND}}-\frac{1}{2}J_{\mathrm{HD}}$ |
| | $\|2,3,1\rangle$ | $\frac{1}{4}J_{\mathrm{NH}}$ | $-2J_{\mathrm{ND}}-2J_{\mathrm{HD}}$ |
| | $\|3,2,1\rangle$ | $\frac{1}{4}J_{\mathrm{NH}}$ | $J_{\mathrm{ND}}+J_{\mathrm{HD}}$ |
| | $\|2,2,1\rangle$ | $\frac{1}{4}J_{\mathrm{NH}}$ | $-\frac{1}{2}J_{\mathrm{ND}}-\frac{1}{2}J_{\mathrm{HD}}$ |
| | $\|1,2,1\rangle$ | $\frac{1}{4}J_{\mathrm{NH}}$ | $-\frac{3}{2}J_{\mathrm{ND}}-\frac{3}{2}J_{\mathrm{HD}}$ |
| | $\|2,1,1\rangle$ | $\frac{1}{4}J_{\mathrm{NH}}$ | $\frac{1}{2}J_{\mathrm{ND}}+\frac{1}{2}J_{\mathrm{HD}}$ |
| | $\|1,1,1\rangle$ | $\frac{1}{4}J_{\mathrm{NH}}$ | $-\frac{1}{2}J_{\mathrm{ND}}-\frac{1}{2}J_{\mathrm{HD}}$ |
| | $\|0,1,1\rangle$ | $\frac{1}{4}J_{\mathrm{NH}}$ | $-J_{\mathrm{ND}}-J_{\mathrm{HD}}$ |
| | $\|1,0,1\rangle$ | $\frac{1}{4}J_{\mathrm{NH}}$ | 0 |
| | $\|3,3,0\rangle$ | $-\frac{3}{4}J_{\mathrm{NH}}$ | 0 |
| | $\|2,2,0\rangle$ | $-\frac{3}{4}J_{\mathrm{NH}}$ | 0 |
| | $\|1,1,0\rangle$ | $-\frac{3}{4}J_{\mathrm{NH}}$ | 0 |
| | $\|0,0,0\rangle$ | $-\frac{3}{4}J_{\mathrm{NH}}$ | 0 |

| Transition | Frequency (high) |
|---|---|
| $\|4,3,1\rangle \rightarrow \|3,3,0\rangle$ | $J_{\mathrm{NH}}+\frac{3}{2}J_{\mathrm{ND}}+\frac{3}{2}J_{\mathrm{HD}}$ |
| $\|3,3,1\rangle \rightarrow \|3,3,0\rangle$ | $J_{\mathrm{NH}} - \frac{1}{2}J_{\mathrm{ND}} - \frac{1}{2}J_{\mathrm{HD}}$ |
| $\|2,3,1\rangle \rightarrow \|3,3,0\rangle$ | $J_{\mathrm{NH}} - 2J_{\mathrm{ND}} - 2J_{\mathrm{HD}}$ |
| $\|3,2,1\rangle \rightarrow \|2,2,0\rangle$ | $J_{\mathrm{NH}} + J_{\mathrm{ND}} + J_{\mathrm{HD}}$ |
| $\|2,2,1\rangle \rightarrow \|2,2,0\rangle$ | $J_{\mathrm{NH}} - \frac{1}{2}J_{\mathrm{ND}} - \frac{1}{2}J_{\mathrm{HD}}$ |
| $\|1,2,1\rangle \rightarrow \|2,2,0\rangle$ | $J_{\mathrm{NH}} - \frac{3}{2}J_{\mathrm{ND}} - \frac{3}{2}J_{\mathrm{HD}}$ |
| $\|2,1,1\rangle \rightarrow \|1,1,0\rangle$ | $J_{\mathrm{NH}} + \frac{1}{2}J_{\mathrm{ND}} + \frac{1}{2}J_{\mathrm{HD}}$ |
| $\|1,1,1\rangle \rightarrow \|1,1,0\rangle$ | $J_{\mathrm{NH}} - \frac{1}{2}J_{\mathrm{ND}} - \frac{1}{2}J_{\mathrm{HD}}$ |
| $\|0,1,1\rangle \rightarrow \|1,1,0\rangle$ | $J_{\mathrm{NH}} - J_{\mathrm{ND}} - J_{\mathrm{HD}}$ |

| Transition | Spectral frequency (high) |
|---|---|
| $\|4,3,1\rangle \rightarrow \|3,3,1\rangle$ | $2J_{\mathrm{ND}} + 2J_{\mathrm{HD}}$ |
| $\|3,3,1\rangle \rightarrow \|2,3,1\rangle$ | $\frac{3}{2}J_{\mathrm{ND}} + \frac{3}{2}J_{\mathrm{HD}}$ |
| $\|3,2,1\rangle \rightarrow \|2,2,1\rangle$ | $\frac{3}{2}J_{\mathrm{ND}} + \frac{3}{2}J_{\mathrm{HD}}$ |
| $\|3,2,1\rangle \rightarrow \|2,2,1\rangle$ | $J_{\mathrm{ND}} + J_{\mathrm{HD}}$ |
| $\|3,2,1\rangle \rightarrow \|2,2,1\rangle$ | $J_{\mathrm{ND}} + J_{\mathrm{HD}}$ |
| $\|3,2,1\rangle \rightarrow \|2,2,1\rangle$ | $\frac{1}{2}J_{\mathrm{ND}} + \frac{1}{2}J_{\mathrm{HD}}$ |

### e. Zero-field NMR spectrum of $^{15}\mathrm{ND}_4^+$

For $^{15}\mathrm{ND}_4^+$ $(x = 4)$ the spectrum may be solved analytically without using perturbation theory. The Hamiltonian is given by

$$\hat{H} = J_{\mathrm{ND}}\left(\hat{\mathbf{S}} \cdot \hat{\mathbf{K}}_{\mathrm{B}}\right). \tag{S11}$$

The selection rules introduced above remain unchanged, but $\hat{\mathbf{K}}_{\mathrm{A}}$ is now replaced by $\hat{\mathbf{K}}_{\mathrm{B}}$. We therefore obtain the states and transitions shown in **Table S10**. States are denoted as $|F, K_{\mathrm{B}}\rangle$, here $F$ is the quantum number corresponding to $\hat{\mathbf{F}} = \hat{\mathbf{S}} + \hat{\mathbf{K}}_{\mathrm{B}}$.

**Table S10.** Nuclear spin eigenstates, corresponding energies, allowed transitions and zero-field NMR spectral frequencies of $^{15}ND_4^+$ ion.

| Eigenstate, $\lvert F, K_B \rangle$ | Energy | Transition | Frequency |
|---|---|---|---|
| $\left\lvert \frac{9}{2}, 4 \right\rangle$ | $2J_{\mathrm{ND}}$ | $\left\lvert \frac{9}{2}, 4 \right\rangle \rightarrow \left\lvert \frac{7}{2}, 4 \right\rangle$ | $\frac{9}{2} J_{\mathrm{ND}}$ |
| $\left\lvert \frac{7}{2}, 4 \right\rangle$ | $-\frac{5}{2} J_{\mathrm{ND}}$ | $\left\lvert \frac{7}{2}, 3 \right\rangle \rightarrow \left\lvert \frac{5}{2}, 3 \right\rangle$ | $\frac{7}{2} J_{\mathrm{ND}}$ |
| $\left\lvert \frac{7}{2}, 3 \right\rangle$ | $\frac{3}{2} J_{\mathrm{ND}}$ | $\left\lvert \frac{5}{2}, 2 \right\rangle \rightarrow \left\lvert \frac{3}{2}, 2 \right\rangle$ | $\frac{5}{2} J_{\mathrm{ND}}$ |
| $\left\lvert \frac{5}{2}, 3 \right\rangle$ | $-2J_{\mathrm{ND}}$ | $\left\lvert \frac{3}{2}, 1 \right\rangle \rightarrow \left\lvert \frac{1}{2}, 1 \right\rangle$ | $\frac{3}{2} J_{\mathrm{ND}}$ |
| $\left\lvert \frac{5}{2}, 2 \right\rangle$ | $J_{\mathrm{ND}}$ | | |
| $\left\lvert \frac{3}{2}, 2 \right\rangle$ | $-\frac{3}{2} J_{\mathrm{ND}}$ | | |
| $\left\lvert \frac{3}{2}, 1 \right\rangle$ | $\frac{1}{2} J_{\mathrm{ND}}$ | | |
| $\left\lvert \frac{1}{2}, 1 \right\rangle$ | $-J_{\mathrm{ND}}$ | | |
| $\left\lvert \frac{1}{2}, 0 \right\rangle$ | $0$ | | |